\documentclass[reqno]{amsart}

\usepackage{booktabs} % For formal tables
\usepackage{IEEEtrantools}
\usepackage{amssymb,latexsym,amsfonts,amsmath}
\usepackage{graphicx}
\usepackage{mathrsfs}
\usepackage{dsfont}

\usepackage{subfigure}

\usepackage{tikz}
\usetikzlibrary{calc,shapes,arrows}
\usepackage{subfig}

\usepackage{algorithm}
\usepackage{algorithmic}

\usepackage{xspace}

\newtheorem{theorem}{Theorem}[section]
\newtheorem{lemma}[theorem]{Lemma}

\newtheorem{problem}[theorem]{Problem}

\newtheorem{corollary}[theorem]{Corollary}

\newtheorem{definition}[theorem]{Definition}

\newtheorem{remark}[theorem]{Remark}
\newtheorem{assumption}{Assumption}
\numberwithin{equation}{section}

\newcommand{\R}{{\mathbb{R}}}

\newcommand{\N}{{\mathbb{N}}}

\newcommand{\trans}{\mathsf{t}}

\newcommand{\Let}{:=}

\newcommand{\PP}{\mathds{P}}

\usepackage[many]{tcolorbox}
\usetikzlibrary{calc}
\tcbuselibrary{skins}

\newtcolorbox{resp}[1][]{%
	enhanced jigsaw,%
	colback=gray!5!white,%
	colframe=gray!80!black,%
	size=small,%
	boxrule=1pt,%
	halign title=flush center,%
	coltitle=black,%
	breakable,%
	drop shadow=black!50!white,%
	attach boxed title to top left={xshift=1cm,yshift=-\tcboxedtitleheight/2,yshifttext=-\tcboxedtitleheight/2},%
	minipage boxed title=3cm,%
	boxed title style={%
		colback=white,%
		size=fbox,%
		boxrule=1pt,%
		boxsep=2pt,%
		underlay={%
			\coordinate (dotA) at ($(interior.west) + (-0.5pt,0)$);
			\coordinate (dotB) at ($(interior.east) + (0.5pt,0)$);
			\begin{scope}[gray!80!black]
				\fill (dotA) circle (2pt);
				\fill (dotB) circle (2pt);
			\end{scope}
		}%
	},%
	#1%
}

\begin{document}

\begin{abstract}
In this work, we propose a compositional framework for the construction of control barrier functions for networks of continuous-time stochastic hybrid systems enforcing complex logic specifications expressed by finite-state automata. The proposed scheme is based on a notion of so-called \emph{pseudo-barrier functions} computed for subsystems, by employing which one can synthesize hybrid controllers for interconnected systems enforcing complex specifications over a finite-time horizon. Particularly, we first leverage sufficient small-gain type conditions to compositionally construct control barrier functions for interconnected systems based on the corresponding pseudo-barrier functions computed for subsystems. Then, using the constructed control barrier functions, we provide probabilistic guarantees on the satisfaction of given complex specifications in a bounded time horizon. In this respect, we decompose the given complex specification to simpler reachability tasks based on automata representing the complements of original finite-state automata. We then provide systematic approaches to solve those simpler reachability tasks by computing corresponding pseudo-barrier functions. Two different systematic techniques are provided based on (i) the sum-of-squares (SOS) optimization program and (ii) counter-example guided inductive synthesis (CEGIS) to search for pseudo-barrier functions of subsystems while synthesizing local controllers. We demonstrate the effectiveness of our proposed results by applying them to a \emph{fully-interconnected} Kuramoto network of $100$ \emph{nonlinear} oscillators with Markovian switching signals.
\end{abstract}

\title[Compositional Construction of Control Barrier Functions for Stochastic Hybrid Systems]{Compositional Construction of Control Barrier Functions for Continuous-Time Stochastic Hybrid Systems}

\author{Ameneh Nejati$^1$}
\author{Sadegh Soudjani$^2$}
\author{Majid Zamani$^{3,4}$}
\address{$^1$Department of Electrical and Computer Engineering, Technical University of Munich, Germany.}
\email{amy.nejati@tum.de}
\address{$^2$School of Computing, Newcastle University, United Kingdom}
\email{sadegh.soudjani@newcastle.ac.uk}
\address{$^3$Department of Computer Science, University of Colorado Boulder, USA}
\address{$^4$Department of Computer Science, LMU Munich, Germany}
\email{majid.zamani@colorado.edu}
\maketitle

\section{Introduction}

{\bf Motivations and State of the Art.}
In the past few years, formal methods have been becoming a promising approach to automatically synthesize controllers for complex stochastic systems against high-level logic properties, \emph{e.g.,} those expressed as linear temporal logic (LTL) formulae~\cite{pnueli1977temporal}, in a reliable way. Since the closed-form characterization of synthesized policies for stochastic systems with both continuous time and space is not available in general, formal policy synthesis for those complex systems is naturally very challenging mainly due to their computational complexity arising from uncountable sets of states and inputs, especially in many safety-critical applications.

To mitigate the encountered computational complexity, one potential solution proposed in the relevant literature is to approximate original models by simpler ones with finite state sets (finite abstractions). However, the major bottleneck of finite-abstraction techniques is their dependency in the state and input set discretization parameters, and consequently, they suffer from the curse of dimensionality: the computational complexity grows exponentially as the dimension of the system increases. To alleviate this issue, compositional techniques have been introduced in the past few years to construct  finite abstractions of interconnected systems based on abstractions of smaller subsystems~\cite{hahn2013compositional,lavaei2019HSCC_J,lavaei2018ADHSJ,lavaei2018CDCJ}.

Although the proposed compositional frameworks in the setting of finite abstractions can mitigate the effects of the state-explosion problem, the curse of dimensionality may still exist even in smaller subsystems given the range of state and input sets. These challenges motivate a need to employ \emph{control barrier functions} as a discretization-free approach for the controller synthesis of complex stochastic systems. In this respect, discretization-free techniques based on barrier functions for stochastic hybrid systems are initially proposed in~\cite{prajna2004safety,prajna2007framework}. Although~\cite{prajna2007framework} studies the same class of Markovian switching as the one we consider here (cf.~\eqref{Markovian}), the proposed results in~\cite{prajna2007framework} only address the verification problem and from a monolithic point of view. In comparison, we deal here with a controller synthesis problem by proposing a compositional approach to handle large-scale stochastic systems. Stochastic safety verification using barrier certificates for switched diffusion processes and stochastic hybrid systems is respectively proposed in~\cite{wisniewski2017stochastic} and~\cite{huang2017probabilistic}. A verification approach for stochastic switched systems via barrier functions is presented in~\cite{Mahathi2019}. 
Verification of Markov decision processes using barrier certificates is proposed in~\cite{M.Ahmadi}. Temporal logic controller synthesis of stochastic systems via control barrier functions is presented in~\cite{Pushpak2019}. 
Compositional construction of control barrier certificates for discrete-time stochastic switched systems is studied in~\cite{Amy_LCSS20}. Although the proposed results in~\cite{Amy_LCSS20} are dealing with stochastic switched systems, their switching signals are deterministic that are served as control inputs. In comparison, switching signals here are not control inputs and are randomly changing. As a result, the controller synthesis problem here is more challenging compared to~\cite{Amy_LCSS20} since it deals with two different types of inputs: (i) internal inputs modeling the effects of other subsystems, and (ii) switching signals which are randomly changing the modes of the system.

Finite-time horizon safety verification of stochastic nonlinear systems with both continuous time and state-space using barrier certificates is proposed in~\cite{steinhardt2012finite}. Control barrier certificates for a class of stochastic nonlinear systems against safety specifications are discussed in~\cite{liu2018adaptive}. A controller synthesis framework  for stochastic control systems based on control barrier functions is also provided in~\cite{clark2019control}. The results in \cite{clark2019control} consider both complete information systems, in which the controller has access to the full-state information, as well as incomplete information systems where the state must be reconstructed from noisy measurements. Verification of uncertain partially-observable Markov decision processes (POMDPs) with uncertain transition and/or observation probabilities using barrier certificates is discussed in~\cite{ahmadi2018verification}. A policy synthesis in multi-agent POMDPs via discrete-time barrier functions to enforce safety specifications is proposed in~\cite{ahmadi2019safe}. An introduction and overview
of recent results on control barrier functions and their applications in verifying and enforcing safety properties is surveyed in~\cite{ames2019control}.

{\bf Contributions.} 
In this paper, we propose a compositional approach for the construction of control barrier functions for continuous-time stochastic hybrid systems enforcing complex logic specifications expressed by finite-state automata. We first compositionally construct control barrier functions for interconnected systems based on so-called pseudo-barrier functions of subsystems by leveraging some sufficient small-gain type conditions. Given the constructed control barrier functions, we then provide probabilistic guarantees on the satisfaction of specifications in a bounded time horizon. To do so, we decompose the given complex specification to reachability tasks based on automata representing the complements of original finite-state automata and provide upper bounds on probabilities of satisfaction for those reachability tasks by computing corresponding pseudo-barrier functions.

We utilize two different systematic techniques based on (i) the sum-of-squares (SOS) optimization program~\cite{parrilo2003semidefinite} and (ii) counter-example guided
inductive synthesis (CEGIS) framework~\cite{ravanbakhsh2015counter} to search for pseudo-barrier functions of subsystems while synthesizing local controllers. While the former is appropriate for continuous inputs, the latter is applicable for dynamical systems with finite input sets as well. To show the applicability of our approach to strongly-connected networks with nonlinear dynamics, we apply our proposed techniques to a \emph{fully-interconnected} Kuramoto network of $100$ \emph{nonlinear} oscillators by compositionally synthesizing hybrid controllers regulating the phase of each oscillator in a comfort zone for a finite-time horizon. Proofs of all statements are provided in Appendix.

{\bf Related Work.}
A limited subset of the provided results in this work has been presented in~\cite{Amy_IFAC20}. Our approach here differs from the one in~\cite{Amy_IFAC20} in four main directions. First and foremost, we generalize the underlying dynamics to stochastic \emph{switching} systems with \emph{random} switching signals and solve the controller synthesis problem for this class of systems with respect to high-level logic properties in a compositional manner. The controller synthesis problem now is more challenging since it deals with two different types of inputs: (i) internal inputs modeling the effects of other subsystems, and (ii) switching signals which are randomly changing. Second, we enlarge the class of specifications to those that can be expressed by the accepting language of deterministic finite automata (DFA), whereas~\cite{Amy_IFAC20} handles only invariance specifications. As our third contribution, we enlarge the class of systems to a class of stochastic \emph{hybrid} ones by adding Poisson processes to the dynamics, while the results in~\cite{Amy_IFAC20} only deal with stochastic control systems. Lastly, we provide here an additional approach to compute pseudo-barrier
functions for systems with finite input sets by employing counter-example
guided inductive synthesis framework based on the satisfiability modulo theories (SMT) solvers such as Z3 \cite{de2008z3}, dReal \cite{gao2012delta} or MathSat \cite{cimatti2013mathsat5}.

Compositional construction of control barrier certificates for networks of control systems is also presented in~\cite{Pushpak2020HSCC}. Our proposed approach differs from the one in~\cite{Pushpak2020HSCC} in three directions. First, the results of~\cite{Pushpak2020HSCC} are presented for discrete-time control systems whereas our proposed approaches are developed for a class of continuous-time hybrid systems with random switching signals. Second, the provided results in~\cite{Pushpak2020HSCC} are valid for non-stochastic settings while our work deals with stochastic ones. As the third distinction, the results in~\cite{Pushpak2020HSCC} are based on a different compositionality condition using $\max$ small-gain approach while we propose here a sum-type small-gain condition in which a spectral radius of some matrix needs to be strictly less than one (cf. Remark~\ref{Rem2}).

\section{Continuous-Time Stochastic Hybrid Systems}\label{ct-SHS}

\subsection{Notation}
We denote sets of nonnegative and positive integers by $\N_0 := \{0, 1, 2, \ldots\}$ and $\N := \{1, 2, 3, \ldots \}$, respectively. Symbols $\mathbb R$, $\mathbb R_{>0}$, and $\mathbb R_{\ge 0}$ denote respectively sets of real, positive, and nonnegative real numbers. We denote by $\mathsf{diag}(a_1,\ldots,a_N)$ and $\mathsf{blkdiag}(a_1,\ldots,a_N)$, respectively, a diagonal matrix in $\R^{N\times{N}}$ with diagonal scalar and matrix entries $a_1,\ldots,a_N$ starting from the upper-left corner. Given a matrix $A\in\mathbb{R}^{n\times m}$, $\mathsf{Tr}(A)$ represents the trace of $A$ which is the sum of all its diagonal elements. Symbols $\mathds{I}_n$, $\mathbf{0}_n$, and $\mathds{1}_n$ denote the identity matrix in $\mathbb R^{n\times{n}}$ and the column vectors in $\mathbb R^{n\times{1}}$ with all elements equal to zero and one, respectively. We denote the spectral radius of a matrix $P\in\mathbb R^{n\times n}$ by $\rho_{spc}(P)$ which is defined as $\rho_{spc}(P) = \max\{|\textsf{eig}_1|, \dots,|\textsf{eig}_n|\}$, where $\textsf{eig}_1, \dots,\textsf{eig}_n$ are eigenvalues of $P$. We employ $x = [x_1;\ldots;x_N]$ to denote the corresponding vector of dimension $\sum_i n_i$, given $N$ vectors $x_i \in \mathbb R^{n_i}$, $n_i\in \mathbb N_{\ge 1}$, and $i\in\{1,\ldots,N\}$. Given a vector $x\in\mathbb{R}^{n}$, $\Vert x\Vert$ denotes the Euclidean norm of $x$. For any set $X$ we denote by $2^X$ the power set of $X$ that is the set of all subsets of $X$.  For any set $X$, $|X|$ and $Int(X)$ represent respectively the cardinality and interior of the set. The empty set is denoted by $\emptyset$. Given a set $X$ and $P \subset X$, the complement of $P$ with respect to $X$ is denoted by $X\backslash P= \{x \,\big|\, x \in X, x \notin P\}$. We denote the disjunction ($\vee$) and conjunction ($\wedge$) of Boolean functions $f:\Gamma\rightarrow \{0,1\}$ over an index set $\Gamma$ by $\underset{\alpha\in\Gamma}\vee f(\alpha)$ and $\underset{\alpha\in\Gamma}\wedge f(\alpha)$, respectively. Given functions $f_i:X_i\rightarrow Y_i$, for any $i\in\{1,\ldots,N\}$, their Cartesian product $\prod_{i=1}^{N}f_i:\prod_{i=1}^{N}X_i\rightarrow\prod_{i=1}^{N}Y_i$ is defined as $(\prod_{i=1}^{N}f_i)(x_1,\ldots,x_N)=[f_1(x_1);\ldots;f_N(x_N)]$.
A function $\gamma:\mathbb\mathbb \mathbb R_{\ge 0}\rightarrow\mathbb\mathbb \mathbb R_{\ge 0}$, is said to be a class $\mathcal{K}$ function if it is continuous, strictly increasing, and $\gamma(0)=0$.
A class $\mathcal{K}$ function $\gamma(\cdot)$ is said to be a class $\mathcal{K}_{\infty}$ if
$\gamma(r) \rightarrow \infty$ as $r\rightarrow\infty$.

\subsection{Preliminaries}
We consider a probability space $(\Omega,\mathcal F_{\Omega},\mathds{P}_{\Omega})$, where $\Omega$ is the sample space,
$\mathcal F_{\Omega}$ is a sigma-algebra on $\Omega$ comprising subsets of $\Omega$ as events, and $\mathds{P}_{\Omega}$ is a probability measure that assigns probabilities to events. We assume that triple $(\Omega,\mathcal F_{\Omega},\mathds{P}_{\Omega})$ is endowed with a filtration $\mathbb{F} = (\mathcal F_s)_{s\geq 0}$ satisfying the usual conditions of completeness and right continuity. Let $(\mathbb W_s)_{s \ge 0}$ be a ${\textsf b}$-dimensional $\mathbb{F}$-Brownian motion, and $(\mathbb P_s)_{s \ge 0}$ be an ${\textsf r}$-dimensional $\mathbb{F}$-Poisson process. We assume that the Poisson process and Brownian motion are independent of each other. The Poisson process $\mathbb P_s = [\mathbb P_s^1; \cdots; \mathbb P_s^{\textsf r}]$ models ${\textsf r}$ events whose occurrences are assumed to be independent of each other.

\subsection{Continuous-Time Stochastic Hybrid Systems}\label{systems1}
We consider continuous-time stochastic hybrid systems (ct-SHS) as formalized in the following definition.

\begin{definition}
	A continuous-time stochastic hybrid system (ct-SHS) in this paper is characterized by the tuple
	\begin{align}\label{eq:ct-SHS}
		\Sigma=(X,U,W, \mathcal U,\mathcal W,P,\mathcal{P},\hat f,\hat\sigma,\hat\rho,Y,h),
	\end{align}
	where:
	\begin{itemize}
		\item $X\subseteq \mathbb R^n$ is the state set of the system;
		\item $U\subseteq \mathbb R^{\bar m}$ is the \emph{external} input set of the system;
		\item $W\subseteq \mathbb R^{\hat p}$ is the \emph{internal} input set of the system;
		\item $\mathcal U$ and $\mathcal W$ are, respectively, subsets of sets of $\mathbb{F}$-progressively measurable processes~(see \cite{karatzas2014brownian} for more details) taking values in $\mathbb R^{\bar m}$ and $\mathbb R^{\hat p}$;  
		\item $P = \{1,\dots, m \}$  is a finite set of modes;
		\item $\mathcal{P}$ is a subset of $\mathcal{S}(\mathbb R_{\ge 0},P)$ which denotes the set of piecewise
		constant functions from $\mathbb R_{\ge 0}$ to $P$, continuous from the
		right and with a finite number of discontinuities on every
		bounded interval of $\mathbb R_{\ge 0}$;
		\item $\hat f = \{f_1,\dots, f_m \}$, $\hat\sigma = \{\sigma_1,\dots, \sigma_m \}$, and $\hat\rho = \{\rho_1,\dots, \rho_m \}$ are, respectively, collections of vector fields, diffusion and reset terms indexed
		by $p$. For all $p\in P$, the vector field $f_p:X\times U \times W\rightarrow X$, and diffusion and reset terms $\sigma_p: \mathbb R^n \rightarrow \mathbb R^{n\times \textsf b}$, $\rho_p: \mathbb R^n \rightarrow \mathbb R^{n\times \textsf r}$ are assumed to be globally Lipschitz continuous;
		\item  $Y\subseteq \mathbb R^{\bar q}$ is the output set of the system;
		\item  $h:X\rightarrow Y$ is the output map.
	\end{itemize}
\end{definition}

A continuous-time stochastic hybrid system $\Sigma$ satisfies
\begin{align}\label{sys1}
	\Sigma\!:\left\{\hspace{-1.2mm}\begin{array}{l}\mathsf{d}\xi(t)=f_{\bold{p}(t)}(\xi(t),\nu(t),w(t))\mathsf{d}t+\sigma_{\bold{p}(t)}(\xi(t))\mathsf{d}\mathbb W_t+\rho_{\bold{p}(t)}(\xi(t))\,\mathsf{d}\mathbb P_t,\\
		\zeta(t)=h(\xi(t)),
	\end{array}\right.
\end{align} 
$\mathds P$-almost surely ($\mathds P$-a.s.) for any $\nu \in \mathcal U$, $w \in \mathcal W$, and switching signal $\bold{p}(t):\mathbb R_{\ge 0} \rightarrow P$. For any $p\in P$, we use $\Sigma_p$ to refer to system~\eqref{sys1} with a constant switching signal $\bold{p}(t) = p$ for all $t\in\mathbb R_{\ge 0}$. We denote the \emph{solution process} and \emph{output trajectory} of $\Sigma_p$ with, respectively, stochastic processes $\xi^p:\Omega \times \mathbb R_{\ge 0}\rightarrow X$ and $\zeta^P:\Omega \times \mathbb R_{\ge 0}\rightarrow Y$. We also employ $\xi^p_{x_0 \nu w}(t)$ to denote the value of the solution process at time $t\in\mathbb R_{\ge 0}$ under input trajectories $\nu$ and $w$, and the switching signal $p$ from an initial condition $\xi^p_{x_0 \nu w}(0)= x_0$ $\mathds P$-a.s., where $x_0$ is a random variable that is $\mathcal F_0$-measurable. We also denote by $\zeta^p_{x_0 \nu w}$ the \emph{output trajectory} corresponding to the \emph{solution process} $\xi^p_{x_0 \nu w}$. Here, we assume that Poisson processes $\mathbb P_s^{\bar z}$, for any $\bar z\in \{1,\dots,{\textsf r}\}$, have rates $\bar \lambda_{\bar z}$.

\begin{remark}
	Note that the underlying dynamic considered in~\eqref{sys1} is a class of stochastic hybrid systems in which drift and diffusion terms model the continuous part and the Poisson process models the discrete jump of the system. In particular, Brownian motions and Poisson processes introduce natively two different sources of uncertainty: (i) a continuous random walk throughout the state space that is governed by Brownian motions, and (ii) a discrete random jump with an exponential distribution that is modelled by Poisson processes. In order to study more general dynamical systems, we added Poisson processes to the underlying dynamics. The direct effect of Poisson processes can be seen in the last term of the infinitesimal generator in~\eqref{infinitesimal generator}.
\end{remark}

Given the ct-SHS in~\eqref{sys1} with $p,p'\in P$, the transition probability between modes at any time instant $t\in\mathbb R_{\ge 0}$ is described using the following Markovian switching:
\begin{align}\label{Markovian}
	\PP\Big\{(p,p'), t + \tilde\delta\Big\}\!=\left\{\hspace{-1mm}\begin{array}{l}\tilde\lambda_{pp'}(\xi^{\bold{p}(t)}(t))~\!\tilde\delta, ~~~~~\quad\quad\quad\quad\! \text{if}~ p\neq p',\\
		1 + \tilde\lambda_{pp}(\xi^{\bold{p}(t)}(t))~\!\tilde\delta, \quad\quad\quad\!\! \text{if}~ p= p',\\
	\end{array}\right.
\end{align}
where $\tilde\delta$ is the time increment and $\tilde\lambda_{pp'}: \mathbb R^n \rightarrow \mathbb R$ is a bounded and Lipschitz continuous function representing transition rates, and, for all $x\in \mathbb R^n$, $\tilde\lambda_{pp'}(x) \geq 0$ if $p \neq p'$, and $\sum_{p'\in P}\tilde\lambda_{pp'}(x) = 0$ for all $p\in P$. The Markovian switching in~\eqref{Markovian} implies that the switching between different modes is governed by a continuous-time Markov chain (CTMC)~\cite{aziz2000model}.

\begin{remark}
		Stochastic hybrid systems~\cite{blom2006stochastic,cassandras2006stochastic}, studied in this work, have broad applications in real-life safety-critical systems such as biological networks~\cite{manes2017modeling,intep2009switching,angius2015approximate}, communication networks~\cite{hespanha2004stochastic}, power grids~\cite{samadi2017stochastic,dabbaghjamanesh2018new}, health and epidemiology~\cite{ogura2015disease,nowzari2016analysis}, air traffic networks~\cite{glover2004stochastic}, and manufacturing systems~\cite{ghosh1993optimal}, to name a few.
\end{remark}

\begin{remark}
	In this work, we assume that the controller has access to switching modes, which is a standard assumption used in the relevant literature~\cite{du2021certainty}. In particular, it is supposed that there is a mode detection device which is capable of identifying the system mode in real time so that the controller can switch to the matched mode. There are some results, in the context of stability analysis of stochastic switching systems~\cite{zhang2019analysis,ren2016stability}, which also consider some delay while deploying the synthesized controllers. However, this issue is out of the scope of this work and we leave it to future works.
\end{remark}

Since the main contribution of this work is to propose a compositional approach for the construction of control barrier functions, we are eventually interested in investigating interconnected systems without having internal inputs. In this case, the tuple~\eqref{eq:ct-SHS} reduces to $(X,U,\mathcal U,P,\mathcal{P},\hat f,\hat\sigma,\hat\rho)$ with $f_p:X\times U\rightarrow X$, and ct-SHS~\eqref{sys1} can be re-written as
\begin{align}\notag
	\Sigma:\mathsf{d}\xi(t)=f_{\bold{p}(t)}(\xi(t),\nu(t))\,\mathsf{d}t+\sigma_{\bold{p}(t)}(\xi(t))\,\mathsf{d}\mathbb W_t+\rho_{\bold{p}(t)}(\xi(t))\,\mathsf{d}\mathbb P_t.
\end{align}
We denote by $\mathds{P}^{x_0}_{\nu}$ the probability measure on the trajectory of interconnected systems starting from the initial condition $x_0$ under the external input $\nu$. Note that although we define ct-SHS in~\eqref{sys1} with outputs, we assume the full-state information is available for interconnected systems (\emph{i.e.,} its output map is identity) for the sake of controller synthesis. In particular, the role of outputs in~\eqref{sys1} is mainly for the sake of interconnecting systems. This will be explained in detail in Section~\ref{Sec_CCBC}. In the next sections, we propose an approach for the compositional construction of control barrier functions for interconnected ct-SHS. To achieve this, we define notions of control pseudo-barrier and barrier functions for ct-SHS and interconnected versions, respectively.

\section{Control Pseudo-Barrier and Barrier Functions}\label{Sec_CBC}

In this section, we first introduce a notion of control pseudo-barrier functions (CPBF) for ct-SHS with both internal and external inputs. We then define a notion of control barrier functions (CBF) for ct-SHS with only external inputs. We leverage the former notion to compositionally construct the latter one for interconnected systems. We then employ the latter notion to quantify upper bounds on the probability that the interconnected system reaches certain unsafe regions in a finite-time horizon via Theorem~\ref{barrier}.

\begin{definition}\label{eq:local barrier}
	Consider a ct-SHS $\Sigma_{p}$, and sets $X_{0}, X_{u}\subseteq X$ as initial and unsafe sets of the system, respectively. A twice differentiable function $\mathcal B_{p}:X\rightarrow\mathbb{R}_{\geq0}$ is called a control pseudo-barrier function (CPBF) for $\Sigma_{p}$ if there exist $\alpha_{p},\kappa_{p} \in \mathcal{K}_{\infty}$, $\rho_{\mathrm{int}{p}}\in\mathcal{K}_\infty\cup \{0\}$, and $\gamma_{p},\lambda_{p}, \psi_{p}\in\R_{\geq 0}$, such that for all $p\in P$,
	\begin{align}\label{eq:LB_c1}
		&\mathcal B_{p}(x)\geq\alpha_{p}(\Vert h(x)\Vert^2), \quad\quad\quad\quad\quad\quad\quad\!\!\!\!\!\!\! \forall x\in X,\\\label{eq:LB_c2}
		&\mathcal B_{p}(x)\leq\gamma_{p}, \quad\quad\quad\quad\quad\quad\quad\quad\quad\quad 	\forall x\in X_{0},\\\label{eq:LB_c3}
		&\mathcal B_{p}(x)\geq \lambda_{p}, \quad\quad\quad\quad\quad\quad\quad\quad\quad\quad \forall x\in X_{u},
	\end{align}
	and $\forall x\in X$, $\exists \nu\in U$, such that $\forall w\in W$,
	\begin{align}\label{eq:B_c4}
		\mathcal{L}\mathcal B_{p}(x) + \sum_{p'=1}^m\tilde\lambda_{pp'}(x)\mathcal B_{p'}(x) \leq - \kappa_{p} (\mathcal B_{p}(x))+\rho_{\mathrm{int}{p}}(\Vert w \Vert^2) + \psi_{p}, 
	\end{align}
	where $\mathcal{L} \mathcal B_{p}$ is the \emph{infinitesimal generator} of the stochastic process acting on the function $\mathcal B_{p}$~\cite{oksendal2013stochastic}, as defined in the next definition.
\end{definition}

\begin{definition}
	Consider a ct-SHS $\Sigma_{p}$. The \emph{infinitesimal generator} $\mathcal{L}$ of the process $\xi^{p}(t)$ of $\Sigma_{p}$ acting on the function $\mathcal B_{p}: X\rightarrow R_{\ge 0}$ is defined as
	\begin{align}\label{infinitesimal generator}
		\mathcal{L}\mathcal B_{p}(x)=~\!&\partial_x \mathcal B_{p}(x)f_{p}(x,\nu,w) + \frac{1}{2}\mathsf{Tr}(\sigma_{p}(x)\sigma_{p}(x)^T\partial_{x,x}\mathcal B_{p}(x))+\sum_{j=1}^{\textsf r}\bar\lambda_{j} (\,\mathcal B_{p}(x+\rho_{p}(x)\textsf e_j^\textsf r)- \mathcal B_{p}(x)),
	\end{align}
	where $\textsf e_j^\textsf r$ denotes an $\textsf r$-dimensional vector with $1$ on the $j$-th entry and $0$ elsewhere.
\end{definition}

\begin{remark}
	Note that the CPBF satisfying conditions~\eqref{eq:LB_c1}-\eqref{eq:B_c4} is not useful on its own to ensure the safety of the system. In particular, CPBF captures the role of internal inputs $w$ via $\rho_{\mathrm{int}{p}}$ in~\eqref{eq:B_c4} as the effect of the interaction between subsystems in the interconnection topology and it is useful for the construction of CBF for the overall network. The safety of the overall network can then be verified only using the constructed CBF.
\end{remark}

\begin{remark}
	Condition~\eqref{eq:LB_c1} is required
	for the satisfaction of small-gain type compositionality conditions in Section~\ref{Sec_CCBC}. Although we assume that the full-state information is available for interconnected systems, we define ct-SHS in~\eqref{sys1} with outputs $y = h(x)$, using which we will introduce the interconnection constraint.
\end{remark}

We now adapt the above notion to the interconnected ct-SHS without internal inputs by simply eliminating all terms related
to $w$. This notion will be utilized in Theorem~\ref{barrier} for quantifying upper bounds on probabilities that systems without internal inputs (\emph{e.g.,} interconnected stochastic systems) reach certain unsafe regions.

\begin{definition}\label{eq:barrier}
	Consider the (interconnected) system $\Sigma=(X,U,\mathcal U,P,\mathcal{P},\hat f,\hat\sigma,\hat\rho)$, and $X_{0}, X_{u}\subseteq X$ as, respectively, initial and unsafe sets of the interconnected system. A function $\mathcal B:X \times P\rightarrow\mathbb{R}_{\geq0}$, that is twice differentiable with respect to $x$, is called a control barrier function (CBF) for $\Sigma$ if, for all $p\in P$,
	\begin{align}\label{eq:B1}
		&\mathcal B(x,p)\leq\gamma, \quad\quad\quad\quad\quad \forall x\in X_0, \\\label{eq:B2}
		&\mathcal B(x,p)\geq\lambda,  \quad\quad\quad\quad\quad \forall x\in X_u,
	\end{align}
	and $\forall x\in X$, $\exists \nu\in U$ such that
	\begin{align}\label{eq:B3}
		\mathcal{L}\mathcal B(x,p)  + \sum_{p'=1}^M\tilde\lambda_{pp'}(x) \mathcal B(x,p')\leq - \kappa(\mathcal B(x,p)) + \psi , 
	\end{align}
	for some $\kappa \in \mathcal{K}_{\infty}$, $\gamma, \lambda,\psi\in\R_{\geq 0}$, with $\lambda > \gamma$, and $M = \Pi_{i=1}^Nm_i$, where $m_i$ is the number of modes for each subsystem $\Sigma_i$ as in~\eqref{eq:ct-SHS}.
\end{definition}

\begin{remark}
	CBFs are Lyapunov-like functions defined over the state space of the system enforcing a set of inequalities on both the function itself (i.e., conditions~\eqref{eq:B1}-\eqref{eq:B2}) and the infinitesimal generator of the stochastic process along the flow (i.e., condition~\eqref{eq:B3}). An appropriate
	level set of a barrier certificate (i.e., $\gamma$ here) separates an unsafe region (i.e., $X_u$ here) from all system trajectories starting from a given set of initial states (i.e., $X_0$ here) with some probability lower bound. Moreover, condition~\eqref{eq:B3} ensures that the CBF is decaying up to a nonnegative constant $\psi$, which captures the magnitude of the stochasticity in the system.
\end{remark}

\begin{remark}
	One requires $\gamma<\lambda$ to have a meaningful probabilistic bound using Theorem~\ref{barrier}; however, we only need this condition for the CBF in Definition~\ref{eq:barrier}. One can readily verify that the probabilistic safety guarantee in Theorem~\ref{barrier} is improved by increasing the distance between $\gamma$ and $\lambda$.
\end{remark}

The next theorem shows the usefulness of CBF to quantify upper bounds on probabilities that (interconnected) systems reach certain unsafe regions.

\begin{theorem}\label{barrier}
	Let $\Sigma=(X,U,\mathcal U,P,\mathcal{P},\hat f,\hat\sigma, \hat\rho)$ be an (interconnected) ct-SHS without internal inputs. Suppose $\mathcal B(x,p)$ is a CBF for $\Sigma$ as in Definition~\ref{eq:barrier}, and there exists a constant $\hat\kappa\in\R_{> 0}$ such that the function $\kappa \in \mathcal{K}_{\infty}$ in \eqref{eq:B3} satisfies $\kappa(s)\geq\hat\kappa s$, $\forall s\in\R_{\geq0}$. Then the probability that the solution process of $\Sigma$ starting from any initial state $\xi^p(0)=x_0\in X_0$ and any initial mode $p_0$ reaches $X_u$ under policy $\nu(\cdot)$ within a time horizon $[0,\mathcal T]\subseteq \mathbb R_{\ge 0}$ is formally quantified as
	\begin{align}\label{Kushner1}
		&\mathds{P}^{x_0}_{\nu}\Big\{\xi^p(t)\in X_u \text{ for some } 0\leq t\leq \mathcal T\mid \xi^p(0)=x_0, p_0\Big\}\leq \delta,\\\notag
		&\delta:=
		\begin{cases}
			1-(1-\frac{\gamma}{\lambda})e^{-\frac{\psi \mathcal T}{\lambda}}, \,\,\,\,\,\,\,& \quad \quad\text{if}~\lambda\geq\frac{\psi}{\hat\kappa},\vspace{2mm}\\
			\frac{\hat\kappa \gamma + (e^{\hat\kappa \mathcal T} - 1) \psi}{\hat\kappa \lambda e^{\hat\kappa \mathcal T} }, & \quad \quad\text{if}~\lambda\leq\frac{\psi}{\hat\kappa}.
		\end{cases}
	\end{align}
\end{theorem}

\begin{remark} 
	Note that since control barrier functions provide only sufficient conditions for synthesizing safety controllers and not necessary
	ones, the initial level-set of CBF, i.e., $\mathcal B(x,p) = \gamma$ which is mode-dependent here, is a subset of the maximal winning set. One can always maximize the volume of the initial level-set of CBF, potentially to be close to the maximal winning set, by increasing the degree of CBF but at the cost of having more computational complexity. Remark that due to having unbounded noises in the stochastic setting, the corresponding safety guarantee in Theorem~\ref{barrier} comes with some probability as opposed to deterministic setting where the safety is guaranteed for all realizations.
\end{remark}

The proposed results in Theorem~\ref{barrier} provide upper bounds on the probability that interconnected systems reach unsafe regions in \emph{finite-time} horizons. We now generalize the proposed results to \emph{infinite-time} horizon, as in the next corollary, provided that constant $\psi = 0$ 
\begin{corollary}\label{Kushner5}
	Let $\Sigma=(X,U,\mathcal U,P,\mathcal{P},\hat f,\hat\sigma, \hat\rho)$ be an interconnected ct-SHS without internal inputs. Suppose $\mathcal B(x,p)$ is a CBF for $\Sigma$ such that $\psi = 0$ in~\eqref{eq:B3}. Then the probability that the solution process of $\Sigma$ starting from any initial state $\xi^p(0)=x_0\in X_0$ and any initial mode $p_0$ reaches $X_u$ under policy $\nu(\cdot)$ within a time horizon $[0,\infty)$ is formally quantified as
	\begin{equation*}
		\mathds{P}^{x_0}_{\nu}\Big\{\xi^p(t)\in X_u \text{ for some } 0\leq t< \infty\mid \xi^p(0)=x_0, p_0\Big\}\leq \frac{\gamma}{\lambda}.
	\end{equation*}
\end{corollary}
The proof is similar to that of Theorem~\ref{barrier} by applying \cite[Theorem 12, Chapter II]{1967stochastic} and is omitted here.
\begin{remark} 
	Note that CBF $\mathcal B(x,p)$ satisfying condition~\eqref{eq:B3} with $\psi = 0$ is non-negative supermartingale~\cite[Chapter I]{1967stochastic}. Although the supermartingale property on $\mathcal B$ allows one to provide probabilistic guarantees for infinite-time horizons via Corollary~\ref{Kushner5}, it is restrictive in the sense that a supermartingale CBF $\mathcal B$ may not generally exist \cite{steinhardt2012finite}. Hence, we employ a more general $c$-martingale type condition in our work that does not require such an assumption at the cost of providing probabilistic guarantees only for finite-time horizons.
\end{remark}

In the next section, we analyze networks of stochastic hybrid subsystems
and show under some conditions one can construct CBF of an interconnected system using CPBF of subsystems.

\section{Compositional Construction of CBF}\label{Sec_CCBC}
In this section, we provide a compositional framework for the construction
of control barrier functions for interconnected systems $\Sigma$. Suppose we are given $N$ stochastic hybrid subsystems $\Sigma_i=(X_i,U_i,W_i,\mathcal U_i,\mathcal W_i,P_i,\mathcal{P}_i,\hat f_i,\\\hat\sigma_i,\hat\rho_i, Y_i, h_i), i\in \{1,\dots,N\}$, where their internal inputs and outputs are partitioned as
\begin{align}\notag
	w_i&=[{w_{i1};\ldots;w_{i(i-1)};w_{i(i+1)};\ldots;w_{iN}}],\\\label{config1}
	y_i&=[{y_{i1};\ldots;y_{iN}}],
\end{align}
and their output spaces and functions are of the form
\begin{equation}
	\label{config2}
	Y_i=\prod_{j=1}^{N}Y_{ij},\quad h_i(x_i)=[{h_{i1}(x_i);\ldots;h_{iN}(x_i)}],
\end{equation}
with $h_{ii}(x_i) = x_i$ (\emph{i.e.,} full-state information of subsystems is available). The outputs $y_{ii} = x_i$ are interpreted as \emph{external} ones, whereas the outputs $y_{ij}$ with $i\neq j$ are \emph{internal} ones which are employed to interconnect these stochastic control subsystems. For the interconnection, if there is a connection from $\Sigma_{j}$ to $\Sigma_i$, we assume that $w_{ij}$ is equal to $y_{ji}$. Otherwise, we put the connecting output function identically zero, \emph{i.e.,} $h_{ji}\equiv 0$. Now we define the interconnected stochastic hybrid system.
\begin{definition}\label{interconnection}
	Consider $N\in\mathbb N_{\geq1}$ stochastic hybrid subsystems $\Sigma_i\!=\!(X_i,U_i,W_i,\mathcal U_i,\mathcal W_i,P_i,\mathcal{P}_i,\hat f_i,\hat\sigma_i,\hat\rho_i, Y_i, h_i)$, $i\in \{1,\dots,N\}$, with the input-output configuration as in \eqref{config1} and \eqref{config2}. The 
	interconnection of  $\Sigma_i$, $\forall i\in \{1,\ldots,N\}$, is the interconnected stochastic hybrid system $\Sigma=(X,U,\mathcal U,P,\mathcal{P},\hat f,\hat\sigma,\hat\rho)$, denoted by
	$\mathcal{I}(\Sigma_1,\ldots,\Sigma_N)$, such that $X:=\prod_{i=1}^{N}X_i$, $U:=\prod_{i=1}^{N}U_i$, $P:=\prod_{i=1}^{N}P_i$, $\mathcal{P}:=\prod_{i=1}^{N}\mathcal{P}_i$, $\hat f:=\prod_{i=1}^{N}\hat f_{i}$,
	$\hat\sigma:=\mathsf{blkdiag}(\hat\sigma_1(x_1),\ldots,\\\hat\sigma_N(x_N))$, and $\hat\rho:=\mathsf{blkdiag}(\hat\rho_1(x_1),\ldots,\hat\rho_N(x_N))$, subject to the following constraint:
	\begin{equation}\label{Inclusion}
		\forall i,j\in \{1,\dots,N\},i\neq j\!: ~~~ w_{ji} = y_{ij}, ~~~ Y_{ij}\subseteq W_{ji}.
	\end{equation}
	
\end{definition}
Assume that for $\Sigma_{i{p}_i}, {p}_i\in \{1,\dots,m_i\}, i\in \{1,\dots,N\}$, there exist CPBF $\mathcal B_{i{p}_i}$ as defined in Definition~\ref{eq:local barrier} with functions $\alpha_{i{p}_i},\kappa_{i{p}_i}\in\mathcal{K}_\infty$, $\rho_{\mathrm{int}{i{p}_i}}\in\mathcal{K}_\infty\cup\{0\}$, and constants $\gamma_{i{p}_i},\lambda_{i{p}_i},\psi_{i{p}_i}\in\R_{\geq 0}$. In order to establish the main compositionality result of the paper, we raise the following small-gain type assumption.

\begin{assumption}\label{Asu:1}
	Assume that for any $i,j\in\{1,\cdots,N\}$, $i\neq j$, there exist $\mathcal{K}_\infty$ functions $\hat\gamma_{i}$ and constants $\hat\lambda_{i{p}_i}\in\R_{>0}$ and $\hat\delta_{ij{p}_j}\in\R_{\geq0}$ such that for any $s\in\R_{\geq0}\!:$
	\begin{align}\label{Eq:19}
		&\kappa_{i{p}_i}(s) \geq\hat\lambda_{i{p}_i}\hat\gamma_i(s),\\
		&h_{ji}\equiv 0\implies \hat\delta_{ij{p}_j}=0,\\
		&h_{ji}\not\equiv 0\implies \rho_{\mathrm{int}_{i{p}_i}}((N-1)\alpha_{j{p}_j}^{-1}(s))\leq\hat\delta_{ij{p}_j}\hat\gamma_j(s),\label{Eq:20}
	\end{align}
	where $\alpha_{j{p}_j}$, $\kappa_{i{p}_i}$, and $\rho_{\mathrm{int}_{i{p}_i}}$,  represent the corresponding  $\mathcal{K}_\infty$ functions related to $\mathcal B_{i{p}_i}$ appearing in Definition~\ref{eq:local barrier}.
\end{assumption}

Before presenting the main compositionality theorem, we define $ \Lambda\Let\mathsf{diag}(\hat\lambda_1,\ldots,\hat\lambda_N)$ with $\hat\lambda_{i} = \min_{p_i\in P_i}\{\hat\lambda_{ip_i}\}$,  $\Delta\Let\{\hat\delta_{ij}\}$ with $\hat\delta_{ij} = \max_{p_i\in P_i}\{\delta_{{ij}p_i}\}$ and $\hat\delta_{ii}=0$, $\forall i\in\{1,\cdots,N\}$, and $\Gamma(s)\Let[\hat\gamma_1(s_1);\ldots;\hat\gamma_N(s_N)]$, where $s=[s_1;\ldots;s_N]$.
In the next theorem, we leverage the small-gain Assumption~\ref{Asu:1} to compute compositionally a control barrier function for the interconnected
system $\Sigma$ as in Definition~\ref{eq:barrier}.
\begin{theorem}\label{Thm:2}
	Consider the interconnected stochastic hybrid system $\Sigma=\mathcal{I}(\Sigma_1,\ldots,\Sigma_N)$ induced by $N\in\N_{\geq1}$ stochastic hybrid subsystems~$\Sigma_i$. Suppose that each mode $\Sigma_{i{p}_i}$ admits a CPBF $\mathcal B_{i{p}_i}$ as defined in Definition~\ref{eq:local barrier} with initial and unsafe sets $X_{0_i}$ and $X_{u_i}$, respectively. If Assumption~\ref{Asu:1} holds and there exists a vector $\mu$ with $\mu_i >0, i\in \{1,\dots,N\}$, such that
	\begin{align}\label{Eq:21}
		&\mu^T(-\Lambda+\Delta)< 0,\\\label{Eq:28}
		\sum_{i=1}^N\mu_i \min_{p_i\in P_i}&\{\lambda_{ip_i}\} > \sum_{i=1}^N\mu_i \max_{p_i\in P_i}\{\gamma_{ip_i}\},
	\end{align}
	then
	\begin{align}\label{Overall B}
		\mathcal B(x,p)\Let\sum_{i=1}^N\mu_i\mathcal B_{i{p_i}}(x_i),
	\end{align}
	with $p =[p_1;\dots;p_N], p_i\in \{1,\dots,m_i\}$, is a CBF for the interconnected system $\Sigma=\mathcal{I}(\Sigma_1,\ldots,\Sigma_N)$ with the initial and unsafe set $X_0:=\prod_{i=1}^{N}X_{0_i}$, $X_u:=\prod_{i=1}^{N}X_{u_i}$, respectively.
\end{theorem}

The proof of Theorem~\ref{Thm:2} is provided in Appendix.

\begin{remark}\label{Rem2}
	Assumption~\ref{Asu:1} is a well-established one in the relevant literature~\cite{ito2009small,dashkovskiy2011small} studying the stability of large-scale interconnected systems via ISS Lyapunov functions of subsystems. We utilize this standard assumption to construct CBF of networks based on CPBF of their subsystems. The compositionality condition $\mu^T(-\Lambda+\Delta)< 0$, constructed from the parameters in Assumption 4.2, is automatically satisfied if the spectral radius of $\Lambda^{-1}\Delta$ is strictly less than one \cite{dashkovskiy2011small}, denoted by $\rho_{spc}(\Lambda^{-1}\Delta) < 1$, which is easy to check. If $\Delta$ is irreducible, $\mu$ can be chosen as the left eigenvector of $-\Lambda+\Delta$ corresponding to the largest eigenvalue, which is real and negative by the Perron-Frobenius theorem~\cite{axelsson1994iterative}.
\end{remark}

\begin{remark}
	Note that $\hat \lambda_{i{p}_i}$ and $\hat\delta_{ij{p}_j}$ in Assumption~\ref{Asu:1} are used to capture, respectively, the gains of each individual subsystem and its interaction with other subsystems in the interconnection topology, i.e., $\kappa_{i{p}_i}, \rho_{\mathrm{int}_{i_{p_i}}}$. Those $\hat\lambda_{i{p}_i}$ and $\hat\delta_{ij{p}_j}$ satisfying conditions~\eqref{Eq:19},\eqref{Eq:20} are then utilized for the construction of $\Lambda$ and $\Delta$, and accordingly, establishing the compositionality condition $\rho_{spc}(\Lambda^{-1}\Delta) < 1$. On the downside, the small-gain type requirements inherently condition the spectral radius of the
	interconnection matrix which, in general, depends on the size
	of the graph and can be violated as the number
	of subsystems grows~\cite{das2004some},~\cite[Remark 6.1]{zamani2017compositionalMurat}.  
\end{remark}

\section{Logic Specifications Expressed as DFA}
\label{DFA}

In this work, we deal with the class of specifications expressed by the accepting language of deterministic finite automata (DFA), as formalized in the following definition.

\begin{definition}
	A deterministic finite automaton (DFA) is a tuple $\mathcal{A}=\{Q_{\ell},q_0,\mathsf{\Sigma}_{\textsf{a}},F_{\textsf{a}},\trans\}$, where $Q_{\ell}$ is a finite set of locations, $q_0\subseteq Q_{\ell}$ is the initial location, $\mathsf{\Sigma}_{\textsf{a}}$ is a finite set (a.k.a., alphabet), $F_{\textsf{a}}\subseteq Q_{\ell}$ is a finite set of accepting locations, and $\trans: Q_{\ell}\times\mathsf{\Sigma}_{\textsf{a}}\rightarrow {Q_{\ell}}$ is a transition function.
\end{definition}

We denote the set of states in the DFA that can be reached from $q$ in the presence of input symbol $\bar\sigma$ by $\trans(q,\bar\sigma)$. A finite word (\emph{a.k.a.,} trace) $(\bar\sigma_0, \bar\sigma_1, \ldots , \bar\sigma_{k-1}) \in \mathsf{\Sigma}_{\textsf{a}}^{k}$ is accepted by the DFA if there exists a finite state run $\mathsf{q} =(q_0, q_1, \ldots , q_k) \in Q_{\ell}^{k+1}$ such that $q_0 \in Q_0$, $q_{i+1} = \trans(q_i, \bar\sigma_i)$ for all $0 \leq i < k$ and $q_{k} \in F_{\textsf{a}}$. Accordingly, we denote the set of all finite words accepted by $\mathcal{A}$, \emph{i.e., the language accepted by the DFA $\mathcal{A}$,} by $\mathbb{L}(\mathcal{A})$. We also denote the set of all successor states of a state $q \in Q_{\ell}$ by $\Delta(q)$. 
The complement of a DFA is as a DFA by simply interchanging accepting and non-accepting states~\cite{baier2008principles}.

In this work, we study specifications represented by accepting languages of DFA $\mathcal{A}$ with symbols defined over a set of atomic propositions $\mathcal{AP}$, \textit{i.e.,} $\mathsf{\Sigma}_{\textsf{a}}=2^\mathcal{AP}$.
Without loss of generality, we work in this paper directly with the set of atomic propositions $\mathcal{AP}$ instead of its power set $2^{\mathcal{AP}}$, \emph{i.e.,} $\mathsf{\Sigma}_{\textsf{a}}=\mathcal{AP}$. We are interested in LTL specifications in \emph{finite time horizons}, in which the logic operators used in the definition of LTL will also come with a bound on the time horizon (cf. the case study).

We now define how solution processes of the interconnected ct-SHS $\Sigma$ over a finite-time horizon $\mathcal T$ are related to specifications given by the
accepting language of DFA $\mathcal{A}$ via a measurable labeling function $\mathsf L: X \rightarrow \mathcal{AP}$.

\begin{definition}\label{sys_trace}
	Consider an interconnected ct-SHS $\Sigma=(X,U,\mathcal U,P,\mathcal{P},\hat f,\hat\sigma,\hat\rho)$ and a specification expressed by DFA $\mathcal{A}=\{Q_{\ell},q_0,\mathsf{\Sigma}_{\textsf{a}},F_{\textsf{a}},\trans\}$. Let $\mathsf L: X \rightarrow \mathcal{AP}$ be a measurable labeling function. A finite sequence $\bar\sigma_\xi=(\bar\sigma_0,\bar\sigma_1,\ldots,\bar\sigma_{k-1})\in \mathcal{AP}^k$ is a finite trace of the solution process $\xi_{a\varrho }$ under the control policy $\varrho$ over a finite-time horizon $ [0, \mathcal T]\subseteq R_{\ge 0}$ if there exists an associated time sequence $t_0, t_1,\ldots, t_{k-1}$ such that $t_0 = 0$, $t_k=\mathcal T$, and for all $j \in (0,1,\ldots,k-1)$, $t_j \in R_{\ge 0}$, the following conditions hold:
	\begin{itemize}
		\item $t_j < t_{j+1}$
		\item $\xi(t_j) \in \mathsf L^{-1} (\bar\sigma_j)$
		\item If $\bar\sigma_j \neq \bar\sigma_{j+1}$, then for some $ t'_j \in [t_j,t_{j+1}]$,
		\begin{align}\notag
			\xi(t) \in
			\begin{cases}
				\mathsf L^{-1} (\bar\sigma_j),\,\,\,\,\,\,\,& \forall t \in (t_j,t'_j),\\
				\mathsf L^{-1} (\bar\sigma_{j+1}),\,\,\,\,\,\,\,& \forall t \in (t'_j,t_{j+1}).
			\end{cases}
		\end{align}
		In other words,
		\begin{align}\notag
			\xi(t'_j) \in
			\mathsf L^{-1} (\bar\sigma_j) ~~ \text{or} ~~\mathsf L^{-1} (\bar\sigma_{j+1}).
		\end{align}
	\end{itemize}
\end{definition}

We now define the probability of satisfaction under which solution processes of the interconnected system $\Sigma$ over a finite-time horizon $\mathcal T$ fulfill a specification expressed by DFA $\mathcal{A}$.

\begin{definition}
	Consider an interconnected ct-SHS $\Sigma=(X,U,\mathcal U,P,\mathcal{P},\hat f,\hat\sigma,\hat\rho)$, a specification given by the accepting language of DFA $\mathcal{A}=\{Q_{\ell}, q_0,\mathsf{\Sigma}_{\textsf{a}},F_{\textsf{a}},\trans\}$, and a labeling function $\mathsf L:X \rightarrow \mathcal{AP}$. Then, $\mathds{P}_\varrho^{x_0}\{\bar\sigma_\xi \models \mathcal{A}\}$ denotes the probability that solution processes $\xi_{a\varrho}$  under the control policy $\varrho$ with initial condition $\xi(0)=x_0$ satisfy the specification expressed by $\mathcal{A}$ over the finite-time horizon $\mathcal T$.
\end{definition}

\begin{remark}\label{Remark: 2}
	Note that the set of atomic propositions $\mathcal{AP}=\{\bar p_0,\bar p_1,\ldots,\bar p_z\}$ and the labeling function $\mathsf L: X \rightarrow \mathcal{AP}$ provide a measurable partition of the state set $X = \cup_{i=1}^z X_i$ as  $X_i:=\mathsf L^{-1}(\bar p_i)$. Without loss of generality, we assume that $X_i\neq \emptyset$ for any $i$, since all the atomic propositions $\bar p_i$ with $\mathsf L^{-1}(\bar p_i) = \emptyset$ can be replaced by $(\neg\textsf{true})$ without affecting the probability of satisfaction.
\end{remark}

Now we state the main problem that we aim to address in this section.

\begin{resp}
	\begin{problem}\label{Prob}
		Consider an interconnected ct-SHS $\Sigma$, a specification expressed by the accepting language of DFA  $\mathcal{A}$ and a labeling function $\mathsf L$. Compute a control policy $\varrho$ such that $\mathds{P}_\varrho^{x_0}\{\bar\sigma_\xi \models \mathcal{A}\} \geq \bar\delta$, $\bar\delta \in [0,1]$, for all $x_0\in X_0$.
	\end{problem}
\end{resp}

\begin{remark}
	Note that the proposed approach here based on control barrier functions may not be able to find a suitable controller satisfying the desired specification even though there may exist one. In fact, barrier functions provide only sufficient conditions for synthesizing controllers and not the necessary ones.
\end{remark}

To find a solution to Problem~\ref{Prob}, we compute a control policy that guarantees $\mathds{P}_\varrho^{x_0}\{\bar\sigma_\xi \models \mathcal{A}^c\} \leq \delta$ for all $x_0\in \mathsf L^{-1}(\bar p_i)$ and some $i\in\{1,2,\ldots,z\}$, where $\mathcal{A}^c =\{Q_{\ell},q_0,\mathsf{\Sigma}_{\textsf{a}},\bar F_{\textsf{a}},\trans\}$ is a DFA which is the complement of DFA $\mathcal{A}$ with $\bar F_{\textsf{a}} = Q_{\ell}\backslash F_{\textsf{a}}$. Then the lower bound $\bar\delta$ for the same control policy can be obviously achieved by $\bar\delta= 1 - \delta$. Here, we propose our solution to Problem~\ref{Prob} by providing a method to decompose the complement of given specification into simple reachability problems. The main target now is to find a suitable CBF as in Definition~\ref{eq:barrier} together with a controller for the interconnected ct-SHS for simpler reachability tasks. However, finding a CBF for large-scale complex systems can be computationally intractable. Consequently, we first search for CPBF and the controller for each subsystem and then leverage compositionality results of Theorem~\ref{Thm:2} to acquire the overall CBF and the controller for the given interconnected system for each reachability task. We eventually combine the probabilities of different reachability problems in order to acquire an \emph{overall} lower bound on the probability under which solution processes of the system satisfy the overall specification.

\subsection{Sequential Reachability Decomposition}
In this subsection, we describe sequential reachability decomposition using which a complex specification expressed by DFA can be decomposed into simple reachability tasks. We follow a similar approach as the one proposed in~\cite[Section 4]{Pushpak2019} but for networks of continuous-time stochastic systems.

For a DFA $\mathcal{A}$ representing the property of interest, we first construct a complement DFA $\mathcal{A}^c$, whose language contains all finite words not included in $\mathbb{L}(\mathcal{A})$. We then specify all accepting state runs of $\mathcal{A}^c$
and denote the set of all finite accepting state runs excluding self-loops by $\mathcal{R}$. The accepting state runs are then partitioned to sets of sequential state runs of length $3$, where each of them describes a reachability task.

Let $|\mathsf{q}|=k+1$ be the length of the accepting state run and $\mathcal{R}$ be the set of all finite accepting state runs excluding self-loops, where
\begin{align*}
	\mathcal{R} :=\{\mathsf{q}=(q_0,q_1,\ldots,q_{k}) \in Q_{\ell}^{k+1} \big|q_k \in \bar F_{\textsf{a}}, q_{i} \neq q_{i+1}, \forall i < k\}.
\end{align*}

Note that computation of $\mathcal{R}$ can be efficiently performed by considering the DFA $\mathcal{A}^c$ as a directed graph $\mathcal{D}=(\mathcal{V},\mathcal{E})$, where $\mathcal{V}=Q_{\ell}$ and $\mathcal{E} \subseteq \mathcal{V} \times \mathcal{V}$ are vertices and edges, respectively, such that $(q,q') \in \mathcal{E}$ if and only if $q' \neq  q$ and there exists $\bar p \in \mathcal{AP}$ such that $\trans(q,\bar p)= q'$. For any $(q,q')\in\mathcal{E}$, the atomic proposition corresponding to the edge $(q,q')$ is denoted by $\bar\sigma(q,q')$. It can be readily verified that a finite path starting at a vertex  $q_0$ and terminating at a vertex that $q_k \in \bar F_{\textsf{a}}$ is an accepting state run $\mathsf{q}$ of $\mathcal{A}^c$ without any self-loop belonging to $\mathcal{R}$. Then one can readily compute $\mathcal{R}$ by employing available algorithms for the graph theory such as variants of depth first search algorithm~\cite{russell2002artificial}.

For each $\bar p\in\mathcal{AP}$, we define a set  $\mathcal{R}^{\bar p}$ as
\begin{equation*}
	\mathcal{R}^{\bar p}:=\{\mathsf{q}=(q_0,q_1,\ldots,q_k)\in\mathcal{R}\mid \bar\sigma(q_0,q_1)={\bar p}\in\mathcal{AP} \}.
\end{equation*}
We now utilize the definition of $\mathcal{P}^{\bar p}(\mathsf{q})$ as 
   \begin{align*}
	\mathcal{P}^{\bar p}(\mathsf{q}):=\{(q_i,q_{i+1},q_{i+2})\,\big| \, 0\leq i\leq k-2\},
	\end{align*}
\!\!which is the set of state runs of length $3$, to characterize our problem as a multiple of reachability problems. We accordingly denote the set of all reachability elements arising from different accepting state run sequences by $\mathcal{P}(\mathcal{A}^c)=\bigcup_{{\bar p}\in\mathcal{AP}}\bigcup_{\mathsf{q}\in\mathcal{R}^{\bar p}}\mathcal{P}^{\bar p}(\mathsf{q})$. Computation of CBF is performed for each individual reachability problem that is obtained from the elements of $\mathcal{P}(\mathcal{A}^c)$.

To give the reader more insight on the sequential reachability decomposition, we present the following running example.
\begin{figure} 
	\begin{center}
		\includegraphics[width=0.37\linewidth]{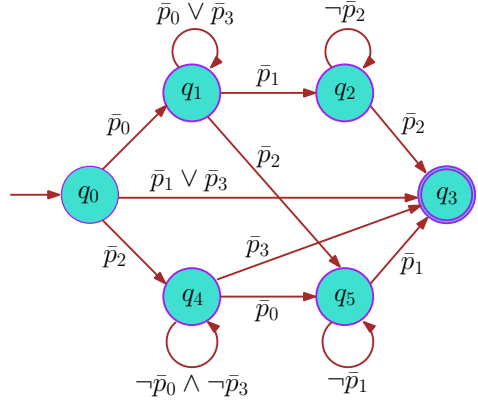} 
		\caption{DFA $\mathcal{A}^c$ in running example.}
		\label{DFA_Exm}
	\end{center}
\end{figure}

{\bf Running Example.}
	Consider the DFA $\mathcal{A}^c$ in Figure~\ref{DFA_Exm} in which $\mathcal{AP}=\{\bar p_0,\bar p_1,\bar p_2,\bar p_3\}$ and $\bar F_{\textsf{a}}=\{q_3\}$. The set of accepting state runs without self-loops is presented as
	\begin{align*}
	\mathcal{R}=\{(q_0,q_1,q_2,q_3), (q_0,q_1,q_5,q_3), (q_0,q_4,q_5,q_3),(q_0, q_4, q_3), (q_0, q_3)\}.
	\end{align*}
	The sets $\mathcal{R}$ for each $\bar p\in\mathcal{AP}$ are defined as:
	\begin{align*}
	&\mathcal{R}^{\bar p_0}= \{(q_0,q_1,q_2,q_3), (q_0,q_1,q_5,q_3) \}, \hspace{1em} \mathcal{R}^{\bar p_1}= \{(q_0, q_3)\} \\
	&\mathcal{R}^{\bar p_2}= \{(q_0,q_4,q_5,q_3), (q_0, q_4, q_3) \}, \hspace{2.3em} \mathcal{R}^{\bar p_3}= \{(q_0, q_3) \}.
	\end{align*} 
	To decompose our complex specification into sequential reachabilities, we consider any $\mathsf{q}\in\mathcal{R}^{\bar p}$ and then define $\mathcal{P}^{\bar p}(\mathsf{q})$ as a set of all state runs of length $3$ as:	
	\begin{align*}
	&\mathcal{P}^{\bar p_0}(q_0,q_1,q_2,q_3)= \{(q_0,q_1,q_2), (q_1,q_2,q_3)\}, \\
	&\mathcal{P}^{\bar p_0}(q_0,q_1,q_5,q_3)=\{(q_0,q_1,q_5),(q_1,q_5,q_3)\},\\
	&\mathcal{P}^{\bar p_2}(q_0,q_4,q_5,q_3)=\{(q_0,q_4,q_5),(q_4,q_5,q_3)\},\\
	&\mathcal{P}^{\bar p_2}(q_0, q_4, q_3)=\{(q_0,q_4,q_3)\}, \\ &\mathcal{P}^{\bar p_1}(q_0,q_3)=\mathcal{P}^{\bar p_3}(q_0,q_3)=\emptyset.
	\end{align*} 
	For every $\mathsf q \in \mathcal{R}^{\bar p}$, the corresponding finite words $\bar\sigma(\mathsf{q})$ are given by
	\begin{align*}
	&\bar\sigma(q_0,q_3)=\{(\bar p_1 \vee \bar p_3)\}, ~~ \bar\sigma(q_0, q_4, q_3)=\{(\bar p_2,\bar p_3)\}, \\
	&\bar\sigma(q_0,q_1,q_2,q_3)=\{(\bar p_0,\bar p_1,\bar p_2)\}, \\ &\bar\sigma(q_0,q_1,q_5,q_3)=\{(\bar p_0,\bar p_2,\bar p_1)\},\\
	&\bar\sigma(q_0,q_4,q_5,q_3)=\{(\bar p_2,\bar p_0,\bar p_1)\}.	
	\end{align*}

The following lemma, as a consequence of Theorem~\ref{barrier},  provides the construction of CBF and its corresponding controller from the elements of $\mathcal{P}^{\bar p}(\mathsf{q})$ constructed from the DFA $\mathcal{A}^c$.   

\begin{lemma}\label{Lemma}
	Consider $(q,q',q'') \in \mathcal{P}^{\bar p}(\mathsf{q})$ for every ${\bar p}\in\mathcal{AP}$ and $\mathsf{q} \in \mathcal{R}^{\bar p}$. The probability that the solution process of ct-SHS $\Sigma$ starting from any initial state $a\in X_0 = \mathsf L^{-1}(\bar\sigma(q,q'))$ under the control policy $\varrho$ reaches $X_u = \mathsf L^{-1}(\bar\sigma(q',q''))$ in a finite-time horizon $\mathcal T$ is upper-bounded by $\delta$ as in~\eqref{Kushner1}, provided that there exists a CBF and a control policy $\varrho$ such that conditions~\eqref{eq:B1}-\eqref{eq:B3} hold.
\end{lemma}

\section{Control Policy}
\label{policy} 

In this section, we follow a similar approach as the one proposed in~\cite[Section 5.1]{Pushpak2019} by combining controllers for reachability tasks to compute a hybrid controller enforcing the overall property.

In our proposed controller synthesis scheme, one needs to compute a CBF and a suitable controller for each element of $\mathcal{P}(\mathcal{A}^c)$. 
Since there may exist two outgoing transitions from a state of the automaton, one should deal with two different controllers which may cause ambiguity while applying them in the closed loop. To tackle this problem, we combine the two reachability problems into one by replacing $X_u$ in Lemma \ref{Lemma} with the union of regions corresponding to the alphabets presenting all the outgoing edges. This technique leads to a \emph{common} CBF and its corresponding controller for different reachability elements in the same partition set.

We partition $\mathcal{P}(\mathcal{A}^c)$ and combine the reachability elements with the same CBF and controller as follows:
\begin{align*}
\bar\gamma_{(q,q',\Delta(q'))} := \{(q,q',q'') \in \mathcal{P}(\mathcal{A}^c)\,|\,  q,q',q''\in Q_{\ell},q'' \in \Delta(q')\}.
\end{align*}

We denote the corresponding CBF and controller to each partition set $\bar\gamma_{(q,q',\Delta(q'))}$ by respectively $\mathcal B_{\bar\gamma_{(q,q',\Delta(q'))}}(x)$ and $\nu_{\bar\gamma_{(q,q',\Delta(q'))}}$. In order to interpret a switching control mechanism, we first define a DFA $\mathcal{A}_\textsf{s}$, which includes the switching mechanism as the following definition.
\begin{definition}
	Consider the DFA $\mathcal{A}^c=\{Q_{\ell},q_0,\mathsf{\Sigma}_{\textsf{a}},\bar F_{\textsf{a}},\trans\}$ with $\bar F_{\textsf{a}} = Q_{\ell}\backslash F_{\textsf{a}}$. The corresponding DFA for switching mechanism is defined as $\mathcal{A}_\textsf{s}=\{Q_{{\ell}_\textsf{s}},q_{0_\textsf{s}},\mathsf{\Sigma}_{\textsf{a}_\textsf{s}},F_{\textsf{a}_\textsf{s}},\trans_\textsf{s}\}$ where
	$Q_{{\ell}_\textsf{s}} := q_{0_\textsf{s}} \cup \{(q,q',\Delta(q')) \, | \, q,q' \in Q_{\ell}\backslash \bar F_{\textsf{a}})\} \cup \bar F_{\textsf{a}}$ is a finite set of locations, $q_{0_\textsf{s}}:= \{(q,\Delta(q)) \, | \, q = q_{0}\}$ is a finite set of initial locations, $\mathsf{\Sigma}_{\textsf{a}_\textsf{s}}=\mathsf{\Sigma}_{\textsf{a}}$ is a finite set as the alphabet for switching mechanism, and  $F_{\textsf{a}_\textsf{s}}=\bar F_{\textsf{a}}$ is a finite set of accepting locations. Moreover, the transition function $\trans_\textsf{s}$ is defined as
	\begin{itemize}
		\item $\forall q_s=(q,\Delta(q))) \in q_{0_\textsf{s}}$, 
		\begin{itemize}
			\item $\trans_\textsf{s}((q,\Delta(q)),\bar\sigma_{(q,q')})=(q,q',\Delta(q'))$ where $q'\in\Delta(q)$,
		\end{itemize}
		\item $\forall q_s=(q,q',\Delta(q')) \in Q_{{\ell}_\textsf{s}} \backslash (q_{0_\textsf{s}} \cup \bar F_{\textsf{a}})$,
		\begin{itemize}
			\item $\trans_\textsf{s}((q,q',\Delta(q')), \bar\sigma_{(q',q'')})=(q',q'',\Delta(q''))$, where $q,q',q'' \in Q_{\ell}, q''\in\Delta(q') \text{ and } q'' \notin \bar F_{\textsf{a}}$,
			\item $\trans_\textsf{s}((q,q',\Delta(q')),\bar\sigma_{(q',q'')})=q''$ where $q,q',q'' \in Q_{\ell}$, $q'' \in \Delta(q')$ and $q'' \in \bar F_{\textsf{a}}$.  
		\end{itemize}
	\end{itemize}
\end{definition}
The switching control mechanism for Problem~\ref{Prob} is formally defined as
\begin{equation*}
\varrho(x,q_\textsf{s})=\nu_{\bar\gamma_{q'_\textsf{s}}}(x), \ \ \forall(q_\textsf{s},\mathsf L(x),q'_\textsf{s}) \in \trans_\textsf{s}.
\end{equation*}

\begin{figure}
	\begin{center}
		\includegraphics[width=0.6\linewidth]{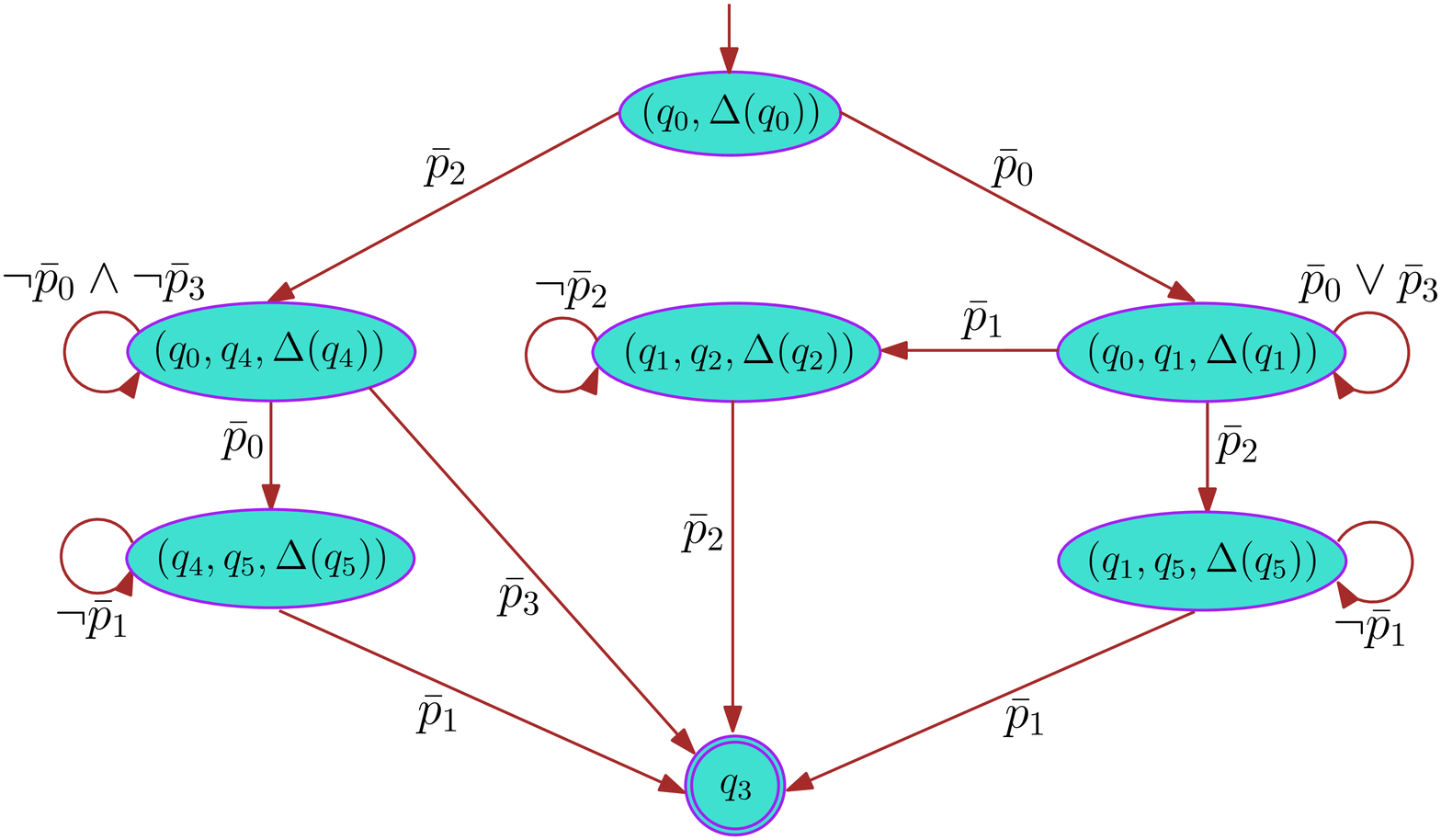} 
		\caption{DFA $\mathcal{A}_\textsf{s}$ describing switching mechanism.}
		\label{DFA_Switching}
	\end{center}
\end{figure}

{\bf Running Example (continued).} Consider the DFA $\mathcal{A}^c$ presented in Figure~\ref{DFA_Exm}. The elements $(q_0,q_1,q_2)$ and $(q_0,q_1,q_5)$ compose two individual reachability problems: one for reaching the region $\mathsf L^{-1}(\bar p_1)$ and the other one for reaching the region $\mathsf L^{-1}(\bar p_2)$ both from the same region $\mathsf L^{-1}(\bar p_0)$. Since there are two outgoing transitions from the same state $q_1$ in the region $\mathsf L^{-1}(\bar p_0)$, there are two different controllers. The DFA $\mathcal{A}_\textsf{s}=\{Q_{{\ell}_\textsf{s}},q_{0_\textsf{s}},\mathsf{\Sigma}_{\textsf{a}_\textsf{s}},F_{\textsf{a}_\textsf{s}},\trans_\textsf{s}\}$ modeling the switching mechanism between different control policies of the DFA $\mathcal{A}^c$ in Figure~\ref{DFA_Exm}, is represented in Figure~\ref{DFA_Switching}.

Now, we propose our solution to compute a lower bound on the probability that the desired specification is satisfied for Problem~\ref{Prob}.

\begin{theorem} \label{th:sumprod}
	Consider a specification expressed by the accepting language of DFA $\mathcal{A}$, and the DFA $\mathcal{A}^c$ as its complement. For  every ${\bar p} \in \mathcal{AP}$, let $\mathcal{R}^{\bar p}$ be the set of all accepting state runs and $\mathcal{P}^{\bar p}(\mathsf{q})$ be the set of state runs of length $3$. Then the probability that the solution process of $\Sigma$ starting from any initial state $\xi(0)=x_0\in \mathsf L^{-1}({\bar p})$ under the corresponding switching control policy satisfying the specification expressed by $\mathcal{A}^c$ over the time horizon $[0,\mathcal T] \subseteq \mathbb R_{\ge 0}$ is upper bounded by
	
	\begin{equation} \label{Sumprod}
		\mathds{P}_\varrho^{x_0}\{\bar\sigma_\xi \models \mathcal{A}^c\} \leq\sum_{q \in \mathcal{R}^{\bar p}} \prod_{\aleph\in\mathcal{P}^{\bar p}(\mathsf{q})}\{\delta_{\aleph} \,\big|\, \aleph=(q,q',q'') \in \mathcal{P}^{\bar p}(\mathsf{q})\}, 
	\end{equation}	
	where $\delta_{\aleph}$ is computed via equation~\eqref{Kushner1} and is the upper bound on the probability that solution processes of $\Sigma$ starting from $X_0 := \mathsf L^{-1}(\bar\sigma(q,q'))$ reach $X_u := \mathsf L^{-1} (\bar\sigma(q',q''))$ within the time horizon $[0,\mathcal T] \subseteq \mathbb R_{\ge 0}$.
\end{theorem}   
Now the probability that solution processes of $\Sigma$ starting from any initial state $\xi(0)=x_0\in \mathsf L^{-1}({\bar p})$ under the same hybrid controller satisfy the specification represented by the language of DFA $\mathcal{A}$ is lower bounded by
\begin{equation}
	\mathds{P}_\varrho^{x_0}\{\bar\sigma_\xi \models \mathcal{A}\} \geq 1-\sum_{q \in \mathcal{R}^{\bar p}}\prod_{\aleph\in\mathcal{P}^{\bar p}(\mathsf{q})}\{\delta_{\aleph} \,\big|\, \aleph=(q,q',q'') \in \mathcal{P}^{\bar p}(\mathsf{q})\}. 
\end{equation}

\section{Computation of CPBF and Controller}\label{compute_CBC}
In  this section, we briefly discuss two different approaches based on sum-of-squares (SOS) optimization and counter-example guided inductive synthesis (CEGIS) to compute CPBF and synthesize corresponding controllers for $\Sigma_p$.
\subsection{Sum-of-Squares Optimization Program}\label{SOSS}
In order to utilize an SOS optimization, we raise the following assumption.
\begin{assumption}\label{ass:BC}
	Suppose that $\Sigma_p$ has a continuous state set $X\subseteq \mathbb R^{n}$ and continuous external and internal input sets $U\subseteq \R^{\bar m}$ and $W\subseteq \R^{\hat p}$. Moreover, the drift term $f_p:X\times U\times W\rightarrow X$ is a polynomial function of the state $x$ and external and internal inputs $\nu, {w}$. Furthermore, the output map $h:X\rightarrow Y$, the diffusion term $\sigma_p: \mathbb R^{n} \rightarrow \mathbb R^{n\times \textsf b}$ and reset term $\rho_p: \mathbb R^{n} \rightarrow \mathbb R^{n\times \textsf r}$ are polynomial functions of the state $x$. We also assume that $\mathcal{K}_{\infty}$ functions $\alpha_p$ and $\rho_{\mathrm{int}p}$ are polynomial.
\end{assumption}
Under Assumption~\ref{ass:BC}, the following lemma provides a set of sufficient conditions for the existence of CPBF required in Definition~\ref{eq:local barrier}, which can be solved as an SOS optimization problem.
\begin{lemma}\label{sos}
	Suppose Assumption~\ref{ass:BC} holds and sets $X_0,X_u, X, U, W$ can be defined by vectors of polynomial inequalities $X_{0}=\{x\in\R^{n}\mid g_{0}(x)\geq0\}$, $X_{u}=\{x\in\R^{n}\mid g_{u}(x)\geq0\}$, $X=\{x\in\R^{n}\mid g(x)\geq0\}$, $U=\{\nu\in\R^{\bar m}\mid g_\nu(\nu)\geq0\}$, and $W=\{w\in\R^{\hat p}\mid g_{w}(w)\geq0\}$, where the inequalities are defined element-wise.
	Suppose there exists an SOS polynomial $\mathcal B_{p}(x)$, constants $\gamma_{p},\psi_{p} \in \R_{\geq 0}$, $\lambda_{p}\in \R_{> 0}$, polynomial functions  $\alpha_{p}, \kappa_{p}\in \mathcal{K}_{\infty}$, $\rho_{\mathrm{int}{p}}\in\mathcal{K}_\infty\cup \{0\}$, polynomials $l_{\nu_{j}}(x)$ corresponding to the $j^{\text{th}}$ input in $\nu=(\nu_1,\nu_2,\ldots,\nu_{\bar m})\in U\subseteq \R^{\bar m}$, and vectors of sum-of-squares polynomials $l_{0_{p}}(x)$, $l_{u_{p}}(x)$, $l_{p}(x)$, $\hat l_{p}(x,\nu,w)$,  $ l_{\nu_{p}}(x,\nu,w)$, and $l_{w_{p}}(x,\nu,w)$ of appropriate dimensions such that the following expressions are sum-of-squares polynomials, for all $p\in P$:
	\begin{align}\label{eq:sos0}
	&\mathcal B_{p}(x)- l_{p}^T(x) g(x)-\alpha_{p}(h(x)^Th(x))\\\label{eq:sos1}
	-&\mathcal B_{p}(x)-l_{0_{p}}^T(x) g_{0}(x)+\gamma_{p}\\\label{eq:sos2}
	&\mathcal B_{p}(x)-l_{u_{p}}^T(x) g_{u}(x)-\lambda_{p} \\\notag
	-&\mathcal{L}\mathcal B_{p}(x) - \sum_{p'=1}^m\tilde\lambda_{pp'}(x)\mathcal B_{p'}(x) - \kappa_{p} (\mathcal B_{p}(x)) +\rho_{\mathrm{int}{p}}(w^Tw)+\psi_{p} - \sum_{j=1}^{\bar m}(\nu_{j}\!-\!l_{\nu_{j_{p}}}(x))\\\label{eq:sos3}
	&-\hat l_{p}^T(x,\nu,w) g(x) - l_{\nu_{p}}^T(x,\nu,w) g_{\nu}(\nu) - l_{w_{p}}^T(x,\nu,w) g_{w}(w).
	\end{align}
	Then, $\mathcal B_{p}(x)$ satisfies conditions~\eqref{eq:LB_c1}-\eqref{eq:B_c4} in Definition~\ref{eq:local barrier} and $\nu=[l_{\nu_{1_{p}}}(x);\dots;l_{\nu_{{\bar m}_{p}}}(x)],$ is the corresponding controller employed at the mode $p\in P$.
\end{lemma}
\begin{remark}
	Note that the function $\kappa_{p}(\cdot)$ in \eqref{eq:sos3} can cause nonlinearity on unknown parameters of $\mathcal{B}_{p}$. A possible way to avoid this issue is to consider a linear function $\kappa_{p}(s)=\hat\kappa_{p} s, \forall s\in\R_{\geq 0}$, with some given constant $\hat\kappa_{p}\in\R_{> 0}$. Then one can employ bisection method to minimize the value of $\hat\kappa_{p}$.
\end{remark}
\begin{remark}
	Note that Lemma~\ref{sos} is different from~\cite[Lemma 5.6]{Pushpak2019} in two main directions. First, there is no such a condition~\eqref{eq:sos0} in~\cite[Lemma 5.6]{Pushpak2019} since this condition here is essential for the sake of interconnection. Besides, we have here internal inputs in the last condition~\eqref{eq:sos3} that model the effects of other subsystems as bounded disturbances.  
\end{remark}

\subsection{Counter-Example Guided Inductive Synthesis}
In this subsection, we search for CPBF for subsystems by employing Satisfiability Modulo Theories (SMT) solvers such as Z3 \cite{de2008z3}, dReal \cite{gao2012delta} or MathSat \cite{cimatti2013mathsat5}. In order to present this framework, we require the following assumption.
\begin{assumption} \label{CEGIS}
	Each $\Sigma_{p}$ has a compact state set $X$, a compact internal input set $W$ and a finite external input set $U$.
\end{assumption}
\begin{remark}
	The assumption of compactness of the state space $X \subseteq \R^n$ can be supported by considering stopped process $\tilde \xi:\Omega \times \mathbb R_{\ge 0}\rightarrow X$ as
	\begin{align}
	&\tilde \xi(t)=
	\begin{cases}
	\xi(t), & \text{for}~t<\tau,\\
	\xi(\tau), & \text{for}~t\geq \tau,
	\end{cases}
	\end{align}
	where $\tau$ is the first time that the solution process $\xi$ of the subsystem exits from the open set $Int(X)$. Note that in most cases, the infinitesimal generator corresponding to $\tilde \xi$ is identical to the one corresponding to $\xi$ over the set $Int(X)$, and is equal to zero outside the set~\cite{1967stochastic}. Hence, the results in Theorem~\ref{barrier} can be employed for any systems with this assumption.
\end{remark}
Now we leverage Assumption~\ref{CEGIS} and reformulate conditions~\eqref{eq:LB_c1}-\eqref{eq:B_c4} as a satisfiability problem as the following lemma.
\begin{lemma}\label{Lemma1}
	Consider a stochastic hybrid subsystem $\Sigma=(X,U,W,\mathcal U,\mathcal W,P,\mathcal{P},\hat f,\hat\sigma,\hat\rho, Y, h),$ fulfilling Assumption \ref{CEGIS}. Suppose there exists a function $\mathcal{B}_{p}(x)$, constants $\gamma_{p},\lambda_{p},\psi_{p} \in \R_{\geq 0}$, and functions  $\alpha_{p}, \kappa_{p}\in \mathcal{K}_{\infty}$, $\rho_{\mathrm{int}{p}}\in\mathcal{K}_\infty\cup \{0\}$, such that the following expression
	is true:
	\begin{align*}
	&\bigwedge_{x \in X} \! \mathcal{B}_{p}(x) \geq \alpha_{p}(\|h(x)\|^2) \!\bigwedge_{x \in X_{0}}  \! \mathcal{B}_{p}(x) \leq \gamma_{p} \nonumber 
	\!\bigwedge_{x \in X_{u}}\! \mathcal{B}_{p}(x) \geq \lambda_{p} \\
	&\bigwedge_{x \in X}(\bigvee_{\nu \in U} (\bigwedge_{w \in W} \!\mathcal{L}\mathcal B_{p}(x)+ \sum_{p'=1}^m\tilde\lambda_{pp'}(x)\mathcal B_{p'}(x)\leq - \kappa_{p} (\mathcal B_{p}(x))+\rho_{\mathrm{int}{p}}(\Vert w \Vert^2) \!+\! \psi_{p})).
	\end{align*}
	Then $\mathcal{B}_{p}(x)$ satisfies conditions~\eqref{eq:LB_c1}-\eqref{eq:B_c4} in Definition~\ref{eq:local barrier}  and accordingly it is a CPBF.
\end{lemma}

\section{Case Study}\label{sec_case}

To show the applicability of our approach to strongly-connected networks with nonlinear dynamics against complex logic properties, we apply our proposed techniques to a \emph{fully-interconnected} Kuramoto network of $100$ \emph{nonlinear} oscillators by compositionally synthesizing hybrid controllers regulating the phase of each oscillator in a comfort zone for a bounded time horizon. Kuramoto oscillator has broad applications in real-life systems such as neural networks, smart grids, automated vehicle coordination, and so on. The model of this case study is adapted from~\cite{skardal2015control} by including stochasticity in
the model. The dynamic for the interconnection of N-oscillators is presented as
\begin{align}\label{Kuramoto}
	\Sigma : \mathsf{d}\theta(t) =(\Omega_{\bold{p}(t)} + \frac{K}{N} \varphi(\theta(t))+\nu(t))\mathsf{d}t+G_{\bold{p}(t)}\mathsf{d}\mathbb W_t + R_{\bold{p}(t)}\mathsf{d}\mathbb P_t,
\end{align}
where $\theta=[\theta_1;\ldots;\theta_N]$ is the phase of oscillators with $\theta_i \in [0,2\pi],\ i=\{1,\ldots,100\}$, $\Omega=[\Omega_1;\ldots;\Omega_N]=\bar\Omega_{p_i}\mathds{1}_N$ is the natural frequency of oscillators with $\bar\Omega_{p_i} =\left\{\hspace{-1.7mm}\begin{array}{l} 0.1,\quad~~\text{if}~~~  p_i = 1,\\
0.12,\quad\text{if}~~~  p_i = 2,\\
\end{array}\right.$ $K=0.001$ is the coupling strength, $\varphi(\theta)=[\varphi(\theta_1);\ldots;\varphi(\theta_N)]$ such that $\varphi(\theta_i)=\Sigma_{j=1}^{N} \text{sin}(\theta_j-\theta_i), i\in\{1,\ldots,100\}$. Moreover, $\nu(t)=[\nu_{1}(t);\ldots;\nu_{N}(t)]$, $G = \bar G_{p_i}\mathds{I}_n$ with $\bar G_{p_i} =\left\{\hspace{-1.7mm}\begin{array}{l} 0.1,\quad~~\text{if}~~~  p_i = 1,\\
0.12,\quad\text{if}~~~  p_i = 2,\\
\end{array}\right.$ and $R = \bar R_{p_i}\mathds{I}_n$ with $\bar R_{p_i} =\left\{\hspace{-1.7mm}\begin{array}{l} 0.1,\quad~~\text{if}~~~  p_i = 1,\\
0.12,\quad\text{if}~~~  p_i = 2.\\
\end{array}\right.$ We consider rates of Poisson processes as $\bar\lambda_{i} = 0.1, \forall i\in\{1,\dots,100\}.$ Now by introducing subsystems $\Sigma_i , i \in \{1,\dots,100\},$ described by
\begin{align}\notag
	\Sigma_i\!:\left\{\hspace{-1mm}\begin{array}{l}{\mathsf{d}\theta_i(t)}=(\bar\Omega_{i\bold{p}_i(t)}+\frac{K}{N}\sum_{j=1,i\neq{j}}^{N}\text{sin}(w_{ij}(t)-\theta_i(t))+\nu_{ip_i}(t))\mathsf{d}t+\bar G_{ip_i}\mathsf{d}\mathbb W_{t_i} + \bar R_{ip_i}\mathsf{d}\mathbb P_{t_i},\\
		\zeta_i(t)=\theta_i(t),\end{array}\right.
\end{align}
one can readily verify that $\Sigma=\mathcal{I}(\Sigma_1,\ldots,\Sigma_N)$ where $w_{ij}(t)=\theta_j(t)$.

\begin{figure}
	\center
	\includegraphics[width=0.35\linewidth]{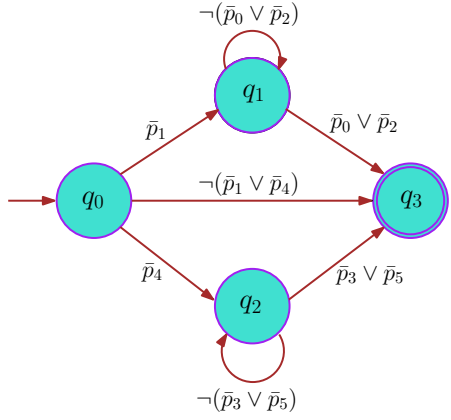}
	\caption{DFA $\mathcal{A}^c$ of the complement of specification.}
	\label{DFA_Kuramoto}
\end{figure}

Transition rates for switching between two modes $P = \{1,2\}$ are given as $\tilde \lambda _{11_i} = -0.9, \tilde \lambda _{12_i} = 0.9, \tilde \lambda _{21_i} = 0.8, \tilde \lambda _{22_i} = -0.8, \forall i\in\{1,\dots,100\}$. In addition, the regions of interest are  $X^0=[0,\frac{\pi}{16}]^{N}, X^1=[\frac{5.8\pi}{12},\frac{6.2\pi}{12}]^N,\\ X^2=[\frac{5.7\pi}{6},\pi]^N, X^3=[\pi,\frac{6.2\pi}{6}]^N, X^4=[\frac{17.8\pi}{12},\frac{18.2\pi}{12}]^N, X^5=[\frac{11.8\pi}{6},2\pi]^N$ and $X^6=X\backslash(X^0 \cup X^1 \cup X^2 \cup X^3 \cup X^4 \cup X^5)$. Each of these regions is associated with atomic propositions given by $\mathcal{AP}=\{\bar p_0,\bar p_1,\bar p_2,\bar p_3,\bar p_4,\bar p_5,\bar p_6\}$ such that the labeling function $\mathsf L(x_z)=\bar p_z$, $\forall x_z \in X^z, z=\{0,\ldots,6\}.$ The objective is to compute a controller such that if the state of the system starts from $X^1$, it always stays away from $X^0$ and $X^2$, and if it starts from $X^4$, it always stays away from $X^3$ and $X^5$ within the time horizon $[0,\mathcal T]$. Such a specification can be represented as an LTL specification given by $(\bar p_1 \wedge \square \neg (\bar p_0 \vee \bar p_2)) \vee (\bar p_4 \wedge \square \neg (\bar p_3 \vee \bar p_5))$ associated with time horizon $\mathcal T = 5$. The specification can also be represented by accepting language $\mathbb{L}(\mathcal{A})$ of a DFA $\mathcal{A}$. Figure~\ref{DFA_Kuramoto} represents the complement DFA $\mathcal{A}^c$.

We decompose the complement of our specification into simple reachability tasks. We consider accepting state runs without self-loops with $\mathcal T=5$. The  DFA $\mathcal{A}^c$ has three accepting state runs $\mathcal{R}=\{(q_0,q_3),(q_0,q_1,q_3),(q_0,q_2,\\q_3)\}$. The sets $\mathcal{P}^{\bar p}(\mathsf{q})$ can be obtained for each of these accepting state runs as $\mathcal{P}^{\bar p_1}(q_0,q_1,q_3)=\{(q_0,q_1,q_3)\}$, and $\mathcal{P}^{\bar p_4}(q_0,q_2,q_3)=\{(q_0,q_2,q_3) \}$. Note that since $(q_0,q_3)$ is a state run of a length $2$ and admits the trivial probability, it is not considered. Accordingly, we need to find control policies and control barrier functions for only two reachability elements. To do so, we utilize the SOS algorithm proposed in~\cite{AmyAutomatica2020} by employing SOSTOOLS and SDP solver SeDuMi. Note that since the dynamics of the system $\Sigma$ in~\eqref{Kuramoto} are non-polynomial and SOS algorithm is only specialized for polynomial dynamics, we first make an approximation to our dynamics. In particular, we take an upper bound on the term $\mathcal{L}\mathcal B_{ip_i}(\theta_i)$ by replacing the $sin(\cdot)$ terms with either $1$ or $-1$.

For the reachability element $(q_0,q_1,q_3)$ with $X_{0_i}=[\frac{5.8\pi}{12},\frac{6.2\pi}{12}]$ and $X_{u_i}=[0,\frac{\pi}{16}] \cup [\frac{5.7\pi}{6},\pi], i\in\{1,\dots,100\}$, we compute CPBF of an order $6$ as $\mathcal B_{ip_i}(\theta_i) = 85\theta_i^6 -310\theta_i^5 + 8.9\theta_i^4 - 36\theta_i^3 + 4791\theta_i^2 - 1038\theta_i + 6245$ and the corresponding hybrid controller $\nu_{ip_i} = - 5356\theta_i + 7000$ for $p_i = 1$; and $\mathcal B_{ip_i}(\theta_i) = 84\theta_i^6 -308\theta_i^5 + 2.8\theta_i^4 - 9.5\theta_i^3 + 4756\theta_i^2 - 1040\theta_i + 6286$ together with $\nu_{ip_i} = - 4229\theta_i + 5000$ for $p_i = 2, \forall i\in\{1,\dots,100\}$. Moreover, the corresponding constants and functions in Definition~\ref{eq:local barrier} satisfying conditions~\eqref{eq:LB_c1}-\eqref{eq:B_c4} are quantified as $\gamma_{ip_i} = 3, \lambda_{ip_i} = 4300, \psi_{ip_i} = 50, \kappa_{ip_i}(s) = 5 \times 10^{-5} s, \alpha_{ip_i} (s) = 0.8 \sqrt{s}, \rho_{\mathrm{int}{ip_i}}(s) = 4 \times10^{-7} \sqrt{s}, \forall s\in\R_{\geq 0}$ for $p_i = 1$; and $\gamma_{ip_i} = 3.2, \lambda_{ip_i} = 4400, \psi_{ip_i} = 52, \kappa_{ip_i}(s) = 53 \times 10^{-6} s, \alpha_{ip_i} (s) = 0.85 \sqrt{s}, \rho_{\mathrm{int}{ip_i}}(s) = 4.2 \times10^{-7} \sqrt{s}, \forall s\in\R_{\geq 0}$ for $p_i = 2, \forall i\in\{1,\dots,100\}$. We now proceed with Theorem~\ref{Thm:2} to construct a CBF for the interconnected system using CPBF of subsystems. One can readily verify that the small-gain Assumption~\ref{Asu:1} holds with $\hat \gamma _i (s) = s, \forall s\in\R_{\geq 0}, \hat \lambda_i = \min_{p_i\in P_i}\{\hat\lambda_{ip_i}\} = 5 \times 10^{-5}$, $\hat \delta_{ij} = \max_{p_i\in P_i}\{\delta_{{ij}p_i}\} = 5\times 10^{-7}$. By selecting $\mu_i = 1, \forall i\in\{1,\dots,100\}$, the spectral radius of $\Lambda^{-1}\Delta$ is computed as $0.99$ which is strictly less that one (cf. Remark~\ref{Rem2}), and consequently, the compositionality condition~\eqref{Eq:21} is satisfied. Moreover, the compositionality condition~\eqref{Eq:28} is also met since $ \min_{p_i\in P_i}\{\lambda_{ip_i}\} > \max_{p_i\in P_i}\{\gamma_{ip_i}\}, \forall i\in \{1,\dots,100\}$. Then by employing the results of Theorem~\ref{Thm:2}, one can conclude that $\mathcal B(\theta,p)\Let\sum_{i=1}^{100}\mathcal B_{i{p_i}}(\theta_i)$ is a CBF for the interconnected system $\Sigma$ with $\gamma = \sum_{i=1}^{100} \max_{p_i\in P_i}\{\gamma_{ip_i}\} = 320, \lambda = \sum_{i=1}^{100}\min_{p_i\in P_i}\{\lambda_{ip_i}\} = 43\times 10^{4}, \kappa(s) = 5\times 10^{-7} s,\forall s\in\R_{\geq 0},$ and $\psi = \sum_{i=1}^{100}\max_{p_i\in P_i}\{\psi_{ip_i}\} = 5200$.

By employing Theorem~\ref{barrier}, one can guarantee that the state of the interconnected system $\Sigma$ starting from the initial set $X_{0} = X^1$ never reaches $X_{u}= X^0 \cup X^2$ during the time horizon $\mathcal T=5$ with the probability of at least $94\%$, \emph{i.e.,}
\begin{equation}\label{prob1}
	\mathds{P}_\varrho^{x_0}\{\bar\sigma_\xi \models \mathcal{A}\} \geq 0.94.
\end{equation}
Similarly, for the reachability element $(q_0,q_2,q_3)$ with $X_{0_i}=[\frac{17.8\pi}{12},\frac{18.2\pi}{12}]$ and $X_{u_i}=[\pi,\frac{6.2\pi}{6}] \cup [\frac{11.8\pi}{6}, 2\pi], i\in\{1,\dots,100\}$, we compute CPBF of an order $6$ as $\mathcal B_{ip_i}(\theta_i) = 0.2\theta_i^6 -0.028\theta_i^5 + 6.7\theta_i^4 - 1.1\theta_i^3 + 20\theta_i^2 - 6365\theta_i + 24559$ and the corresponding hybrid controller $\nu_{ip_i} = - 1733\theta_i + 6900$ for $p_i = 1$; and $\mathcal B_{ip_i}(\theta_i) = 0.11\theta_i^6 -0.038\theta_i^5 + 8.7\theta_i^4 - 5.5\theta_i^3 + 21\theta_i^2 - 5801\theta_i + 22215$ together with $\nu_{ip_i} = - 1678\theta_i + 21870$ for $p_i = 2, \forall i\in\{1,\dots,100\}$. Moreover, the corresponding constants and functions in Definition~\ref{eq:local barrier} satisfying conditions~\eqref{eq:LB_c1}-\eqref{eq:B_c4} are synthesized as $\gamma_{ip_i} = 300, \lambda_{ip_i} = 5000, \psi_{ip_i} = 64, \kappa_{ip_i}(s) = 5 \times 10^{-5} s, \alpha_{ip_i} (s) = 0.8 \sqrt{s}, \rho_{\mathrm{int}{ip_i}}(s) = 4 \times10^{-7} \sqrt{s}, \forall s\in\R_{\geq 0}$ for $p_i = 1$; and $\gamma_{ip_i} = 340, \lambda_{ip_i} = 4500, \psi_{ip_i} = 66, \kappa_{ip_i}(s) = 51 \times 10^{-6} s, \alpha_{ip_i} (s) = 0.82 \sqrt{s}, \rho_{\mathrm{int}{ip_i}}(s) = 4.1 \times10^{-7} \sqrt{s}, \forall s\in\R_{\geq 0}$ for $p_i = 2, \forall i\in\{1,\dots,100\}$. We now proceed with Theorem~\ref{Thm:2} to construct a CBF for the interconnected system using CPBF of subsystems. One can readily verify that the small-gain Assumption~\ref{Asu:1} holds with $\hat \gamma _i (s) = s, \forall s\in\R_{\geq 0}, \hat \lambda_i = \min_{p_i\in P_i}\{\hat\lambda_{ip_i}\} = 5 \times 10^{-5}$, $\hat \delta_{ij} = \max_{p_i\in P_i}\{\delta_{{ij}p_i}\} = 5\times 10^{-7}$. By selecting $\mu_i = 1, \forall i\in\{1,\dots,100\}$, the spectral radius of $\Lambda^{-1}\Delta$ is computed as $0.99$ which is strictly less that one, and consequently, the compositionality condition~\eqref{Eq:21} is satisfied. Moreover, the compositionality condition~\eqref{Eq:28} is also met since $ \min_{p_i\in P_i}\{\lambda_{ip_i}\} > \max_{p_i\in P_i}\{\gamma_{ip_i}\}, \forall i\in \{1,\dots,100\}$. Then by employing the results of Theorem~\ref{Thm:2}, one can conclude that $\mathcal B(\theta,p)\Let\sum_{i=1}^{100}\mathcal B_{i{p_i}}(\theta_i)$  is a CBF for the interconnected system $\Sigma$ with $\gamma = \sum_{i=1}^{100} \max_{p_i\in P_i}\{\gamma_{ip_i}\} = 34000, \lambda = \sum_{i=1}^{100}\min_{p_i\in P_i}\{\lambda_{ip_i}\} = 45\times 10^{4}, \kappa(s) = 5\times 10^{-7} s,\forall s\in\R_{\geq 0},$ and $\psi = \sum_{i=1}^{100}\max_{p_i\in P_i}\{\psi_{ip_i}\} = 6400$.

\begin{figure}
	\center
	\subfigure[]{\includegraphics[width=0.47\linewidth]{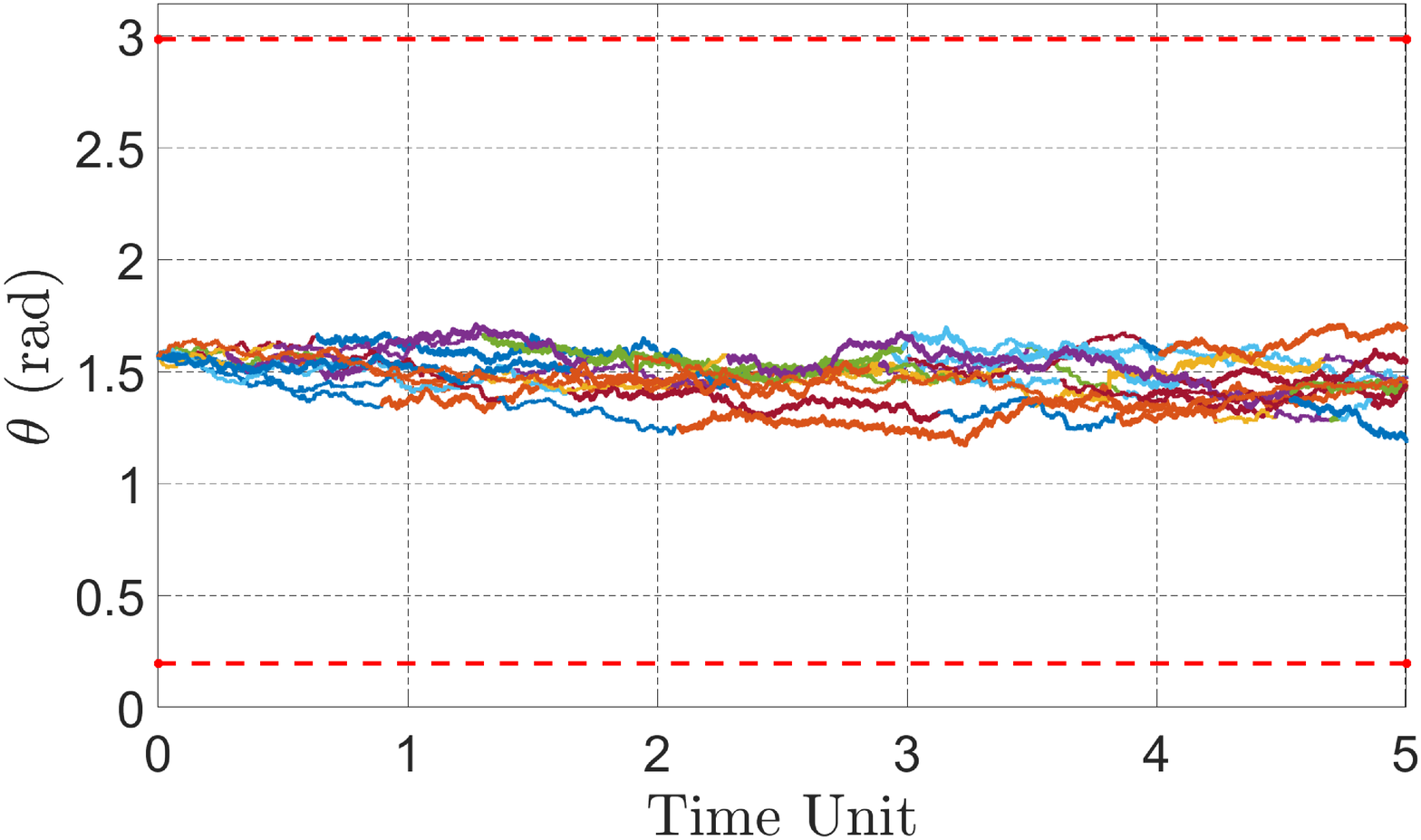}}  \hspace{-0.5cm}
	\subfigure[]{\includegraphics[width=0.47\linewidth]{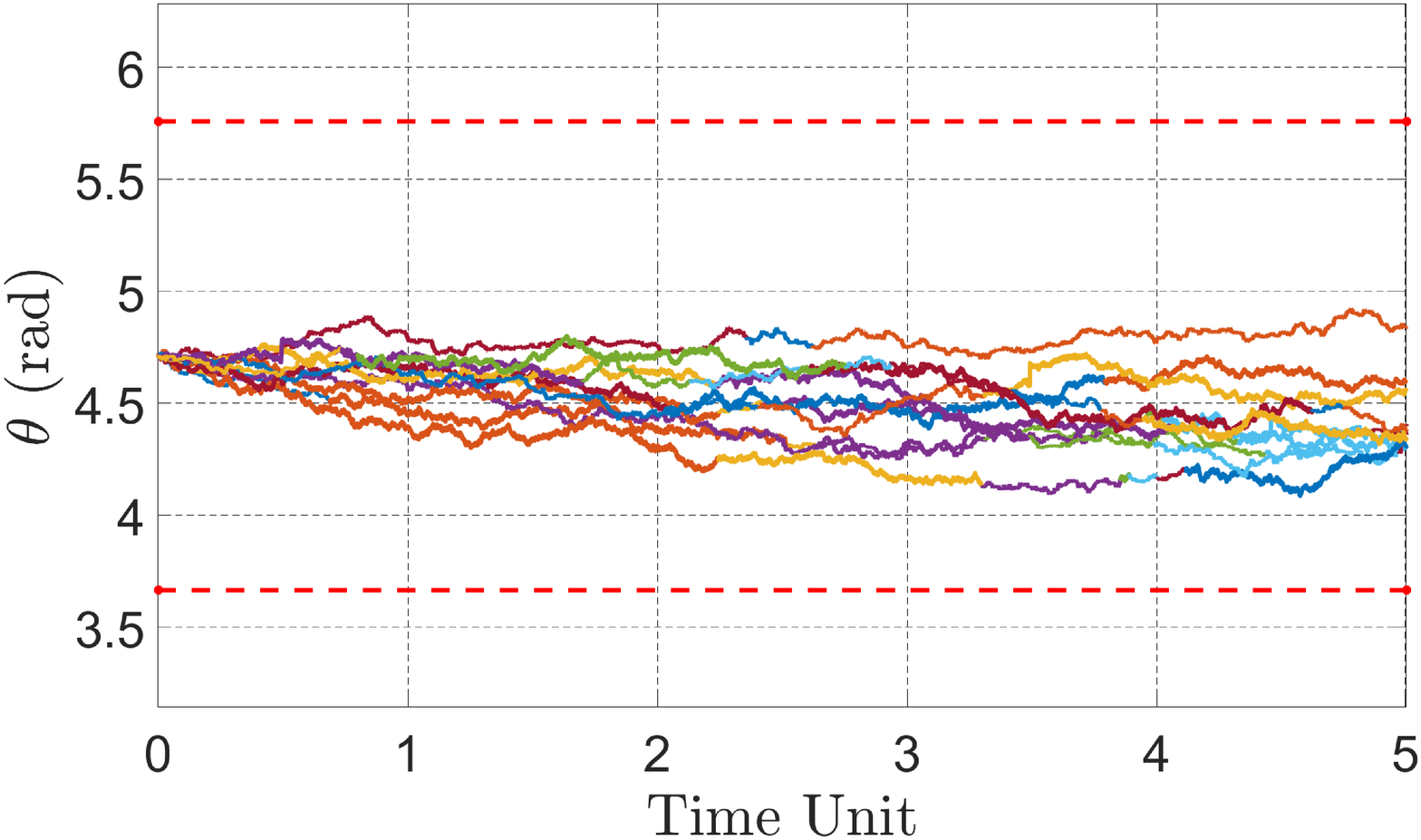}} \vspace{-0.2cm}
	\caption{Closed-loop state trajectories of a representative oscillator with $10$ different noise realizations in a network of $100$ oscillators with initial state starting from (a) Region $X^1$, and (b) Region $X^4$. Changed colours in each trajectory show that the mode is switched to the other one.}
	\label{Kuramoto_Spec}
\end{figure}

By employing Theorem~\ref{barrier}, one can guarantee that the state of the interconnected system $\Sigma$ starting from the initial set $X_{0} = X^4$ never reaches $X_{u}= X^3 \cup X^5$ during the time horizon $\mathcal T=5$ with the probability of at least $86\%$, \emph{i.e.,}
\begin{equation}\label{prob2}
	\mathds{P}_\varrho^{x_0}\{\bar\sigma_\xi \models \mathcal{A}\} \geq 0.86.
\end{equation}
\begin{figure}
	\center
	\includegraphics[width=0.5\linewidth]{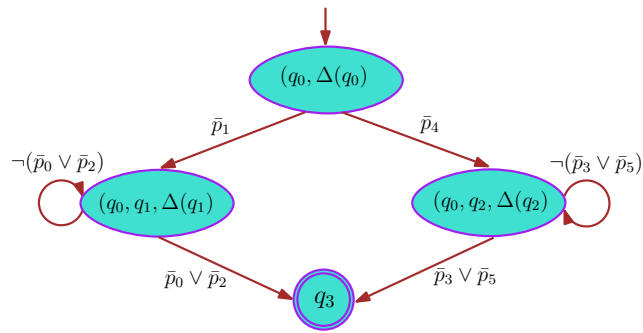}\vspace{-0.2cm}
	\caption{Switching mechanism for controllers.}
	\label{Kuramoto_Switching_Mechanism}
\end{figure}
The switching mechanism for controllers is shown in Figure~\ref{Kuramoto_Switching_Mechanism}. Closed-loop state trajectories of a representative oscillator with $10$ different noise realizations starting from initial regions $X^1$ and $X^4$ are, respectively, illustrated in Figure~\ref{Kuramoto_Spec}-(a) and~\ref{Kuramoto_Spec}-(b). The required computation time and memory usage  for computing CPBF and its corresponding controller for the reachability element $(q_0,q_1,q_3)$  are, respectively, $1.8$ minutes and $23$ MB, and for the reachability element $(q_0,q_2,q_3)$ are, respectively, $1.7$ minutes and $21$ MB on the same machine as the first case study. Note that if one employs our designed controllers and run Monte Carlo simulations on top of the closed-loop system, the empirical probabilities are better than the ones we proposed in~\eqref{prob1},~\eqref{prob2}. However, this issue is expected and the reason is due to the conservatism nature of using polynomial barrier functions with a fix degree, but with the gain of providing a formal lower bound on the probability of satisfaction for safety specification rather than an empirical one.

\section{Conclusion}
In this work, we proposed a compositional scheme for constructing control
barrier functions for continuous-time stochastic hybrid systems to enforce complex logic specifications expressed by finite-state automata. We constructed control barrier functions for large-scale stochastic systems using control pseudo-barrier functions of subsystems as long as some sufficient small-gain type conditions hold. In particular, we decomposed the given complex specification to simpler reachability tasks based on automata representing the complements of original finite-state automata. We then provided a systematic approach to solve those simpler reachability tasks by computing corresponding pseudo-barrier functions. We employed two different techniques based on the sum-of-squares optimization program and counter-example guided inductive synthesis to search for pseudo-barrier functions of subsystems while synthesizing local hybrid controllers. We demonstrated the effectiveness of our proposed results by applying them to a fully-interconnected Kuramoto network of $100$ nonlinear oscillators with Markovian switching signals. Improving the probability of satisfaction for each individual reachability task can be considered as a future research direction.

\section{Acknowledgment}
The authors would like to thank Abolfazl Lavaei for the fruitful discussions and helpful comments.

\bibliographystyle{alpha}
\bibliography{biblio}

\section{Appendix}

\begin{proof}\textbf{(Theorem~\ref{barrier})}
	Based on the condition~\eqref{eq:B2}, we have $X_u\subseteq \{x\in X \,\,\big|\,\, \mathcal B(x) \ge \lambda \}$. Then one has 
	\begin{align}\notag
	&\mathds{P}^{x_0}_{\nu}\Big\{\xi^p(t)\in X_u \text{ for some } 0\leq t\leq \mathcal T\,\,\big|\,\, \xi^p(0)=x_0, p_0\Big\}\\\label{Eq:5}
	&\leq\mathds{P}^{x_0}_{\nu}\Big\{\sup_{0\leq t\leq \mathcal T} \mathcal B(\xi(t))\geq \lambda \,\,\big|\,\, \xi^p(0)=x_0, p_0 \Big\}.
	\end{align}
	One can acquire the upper bound in~\eqref{Kushner1} by applying~\cite[Theorem 1, Chapter III]{1967stochastic} to~\eqref{Eq:5} and utilizing respectively conditions~\eqref{eq:B3} and~\eqref{eq:B1}.
\end{proof}

\begin{proof}\textbf{(Theorem~\ref{Thm:2})}
	We first show that conditions~\eqref{eq:B1} and \eqref{eq:B2} in Definition \ref{eq:barrier} hold. For any $x\Let[x_{1};\ldots;x_{N}] \in X_0 = \prod_{i=1}^{N} X_{0_i} $ and from \eqref{eq:LB_c2}
	\begin{align}\notag
		\mathcal B(x,p)=\sum_{i=1}^N\mu_i\mathcal B_{i{p_i}}(x_i)\leq \sum_{i=1}^N\mu_i \gamma_{ip_i} \leq \sum_{i=1}^N\mu_i \max_{p_i\in P_i}\{\gamma_{ip_i}\} = \gamma,
	\end{align} 
	and similarly for any $x\Let[x_{1};\ldots;x_{N}] \in X_u = \prod_{i=1}^{N} X_{u_i} $ and from \eqref{eq:LB_c3}
	\begin{align*}
		\mathcal B(x,p)=\sum_{i=1}^N\mu_i\mathcal B_{i{p_i}}(x_i)\geq \sum_{i=1}^N\mu_i \lambda_{ip_i} \geq \sum_{i=1}^N\mu_i \min_{p_i\in P_i}\{\lambda_{ip_i}\} = \lambda,
	\end{align*}  
	satisfying conditions \eqref{eq:B1} and \eqref{eq:B2} with $\gamma = \sum_{i=1}^N\mu_i \max_{p_i\in P_i}\{\gamma_{ip_i}\}$ and $\lambda = \sum_{i=1}^N\mu_i \min_{p_i\in P_i}\{\lambda_{ip_i}\}$. Note that $\lambda > \gamma$ according to~\eqref{Eq:28}. Now, we show that condition~\eqref{eq:B3} holds, as well. We first compute $\sum_{p'=1}^M\tilde\lambda_{pp'}(x) \mathcal B(x,p')$ in~\eqref{eq:B3} based on transition rates of subsystems.
	To do so, we first compute it for two subsystems with three modes and then extend it to the  general case of $N$ subsystems with $M$ modes. Consider two subsystems $\Sigma_1,\Sigma_2$ with $3$ independent modes, \emph{i.e.,} $m_1 = 3, m_2 = 3$. The generator matrix~\cite{aziz2000model} of $\Sigma_1,\Sigma_2$ are constructed as
	\begin{align*}
		\tilde Q_{\Sigma_1} \!=\! \begin{bmatrix}
			-\tilde\lambda_{{12}_1} -\tilde\lambda_{{13}_1}& \tilde\lambda_{{12}_1} & \tilde\lambda_{{13}_1}\\
			\tilde\lambda_{{21}_1} & -\tilde\lambda_{{21}_1} -\tilde\lambda_{{23}_1} & \tilde\lambda_{{23}_1}\\
			\tilde\lambda_{{31}_1} & \tilde\lambda_{{32}_1} & -\tilde\lambda_{{31}_1} -\tilde\lambda_{{32}_1}\\
		\end{bmatrix}\!\!,\\
		\tilde Q_{\Sigma_2} \!=\! \begin{bmatrix}
			-\tilde\lambda_{{12}_2} -\tilde\lambda_{{13}_2}& \tilde\lambda_{{12}_2} & \tilde\lambda_{{13}_2}\\
			\tilde\lambda_{{21}_2} & -\tilde\lambda_{{21}_2} -\tilde\lambda_{{23}_2} & \tilde\lambda_{{23}_2}\\
			\tilde\lambda_{{31}_2} & \tilde\lambda_{{32}_2} & -\tilde\lambda_{{31}_2} -\tilde\lambda_{{32}_2}\\
		\end{bmatrix}\!\!.
	\end{align*}
	Now we construct the generator matrix for the interconnected system $\Sigma = \mathcal I(\Sigma_1,\Sigma_2)$ via Table~\ref{Table1}, in which the first and second elements of the pair $(\cdot,\cdot)$ are corresponding to switching modes of the first and second subsystems, respectively. Moreover, diagonal elements ``$*$" are the summation of off-diagonals in each row with a negative sign, \emph{i.e.,} summation of each row including ``$*$" should be zero. For each state $(p_{i_1},p_{j_2})$, we can jump to $(p_{i'_1},p_{j'_2})$ with $ i = i'$ or $j = j'$, \emph{i.e.,} only one of the modes can change~\cite{aziz2000model}.
	
	\begin{table}[ht]
		\small
		\centering
		\caption{\small Generator matrix of the interconnected system $\Sigma = \mathcal I(\Sigma_1,\Sigma_2)$.}
		\vspace{2mm}
		\begin{tabular}{ c |p{0.5cm} | p{0.5cm} | p{0.5cm} | p{0.5cm} | p{0.5cm} | p{0.5cm} | p{0.5cm} | p{0.5cm} | p{0.5cm} |}
			$\!\!\!\tilde Q_{\mathcal I(\Sigma_1,\Sigma_2)}$ & 
			$\!\!\!(1,1)$ &
			$\!\!\!(1,2)$ &
			$\!\!\!(1,3)$ &
			$\!\!\!(2,1)$ &
			$\!\!\!(2,2)$ &
			$\!\!\!(2,3)$ &
			$\!\!\!(3,1)$ &
			$\!\!\!(3,2)$ &
			$\!\!\!(3,3)$
			\\ \hline
			
			$(1,1)$ 
			& * 
			& $\tilde\lambda_{{12}_2}$ 
			& $\tilde\lambda_{{13}_2}$ 
			& $\tilde\lambda_{{12}_1}$ 
			& $0$ 
			& $0$ 
			& $\tilde\lambda_{{13}_1}$ 
			& $0$ 
			& $0$ 
			\\ \hline																
			
			$(1,2)$ 
			& $\tilde\lambda_{{21}_2}$ 
			& * 
			& $\tilde\lambda_{{23}_2}$ 
			& $0$ 
			& $\tilde\lambda_{{12}_1}$ 
			& $0$ 
			& $0$ 
			& $\tilde\lambda_{{13}_1}$
			& $0$
			\\ \hline

			$(1,3)$ 
			& $\tilde\lambda_{{31}_2}$ 
			& $\tilde\lambda_{{32}_2}$ 
			& * 
			& $0$ 
			& $0$ 
			& $\tilde\lambda_{{12}_1}$ 
			& $0$ 
			& $0$ 
			& $\tilde\lambda_{{13}_1}$
			\\ \hline
			
			$(2,1)$ 
			& $\tilde\lambda_{{21}_1}$ 
			& $0$ 
			& $0$ 
			& $*$ 
			& $\tilde\lambda_{{12}_2}$ 
			& $\tilde\lambda_{{13}_2}$ 
			& $\tilde\lambda_{{23}_1}$ 
			& $0$ 
			& $0$
			\\ \hline							
			
			$(2,2)$ 
			& $0$ 
			& $\tilde\lambda_{{21}_1}$ 
			& $0$ 
			& $\tilde\lambda_{{21}_2}$ 
			& $*$ 
			& $\tilde\lambda_{{23}_2}$ 
			& $0$ 
			& $\tilde\lambda_{{23}_1}$ 
			& $0$
			\\ \hline										
			
			$(2,3)$ 
			& $0$ 
			& $0$ 
			& $\tilde\lambda_{{21}_1}$ 
			& $\tilde\lambda_{{31}_2}$ 
			& $\tilde\lambda_{{32}_2}$ 
			& * 
			& $0$ 
			& $0$ 
			& $\tilde\lambda_{{23}_1}$
			\\ \hline	
			
			$(3,1)$ 
			& $\tilde\lambda_{{31}_1}$ 
			& $0$ 
			& $0$ 
			& $\tilde\lambda_{{32}_1}$ 
			& $0$ 
			& $0$ 
			& * 
			& $\tilde\lambda_{{12}_2}$ 
			& $\tilde\lambda_{{13}_2}$
			\\ \hline
			
			$(3,2)$ 
			& $0$ 
			& $\tilde\lambda_{{21}_1}$ 
			& $0$ 
			& $0$ 
			& $\tilde\lambda_{{32}_1}$ 
			& $0$ 
			& $\tilde\lambda_{{21}_2}$ 
			& $*$ 
			& $\tilde\lambda_{{23}_2}$
			\\ \hline
			
			$(3,3)$ 
			& $0$ 
			& $0$ 
			& $\tilde\lambda_{{31}_1}$ 
			& $0$ 
			& $0$ 
			& $\tilde\lambda_{{32}_1}$ 
			& $\tilde\lambda_{{31}_2}$ 
			& $\tilde\lambda_{{32}_2}$ 
			& *
			\\ \hline									
			
		\end{tabular}\vspace{0.2cm}
		\label{Table1}
	\end{table}
	
	Then by employing the definition of CBF in~\eqref{Overall B} and Table~\ref{Table1}, one has 
	\begin{align*}
		&\sum_{p'=1}^9\tilde\lambda_{pp'}(x) \mathcal B(x,p') = \sum_{p'_1=1, p'_2=1}^3\tilde\lambda_{p_1,p_2,p'_1,p'_2}(x) \sum_{i=1}^2\mu_i\mathcal B_{{p'_i}}(x_i)\\
		&=\sum_{p'_1=1, p'_2=1}^3\tilde\lambda_{p_1,p_2,p'_1,p'_2}(x) \mu_1\mathcal B_{1{p'_1}}(x_1) + \sum_{p'_1=1, p'_2=1}^3\tilde\lambda_{p_1,p_2,p'_1,p'_2}(x) \mu_2\mathcal B_{2{p'_2}}(x_2)\\
		& = \sum_{p'_1=1}^3\mu_1\mathcal B_{1{p'_1}}(x_1)  \overbrace{\sum_{p'_2=1}^3\tilde\lambda_{p_1,p_2,p'_1,p'_2}(x)}^{\tilde\lambda_{p_1p'_1}} + \sum_{p'_2=1}^3\mu_2\mathcal B_{2{p'_2}}(x_2) \overbrace{\sum_{p'_1=1}^3\tilde\lambda_{p_1,p_2,p'_1,p'_2}(x)}^{\tilde\lambda_{p_2p'_2}}\\
		&=\sum_{p'_1=1}^3\mu_1\tilde\lambda_{p_1p'_1}\mathcal B_{1{p'_1}}(x_1) + \sum_{p'_2=1}^3\mu_2\tilde\lambda_{p_2p'_2}\mathcal B_{2{p'_2}}(x_2) \\
		&= \sum_{i=1}^2\sum_{p'_i=1}^3 \mu_i \tilde\lambda_{p'_ip_i}(x_i)\mathcal B_{ip'_i}(x_i).
	\end{align*} 
	Note that $\tilde\lambda_{p_1,p_2,p'_1,p'_2}(x)$ is corresponding to the current mode of subsystems 1,2, which is respectively $p_1,p_2$, and the next mode which is $p'_1,p'_2$. One can readily extend the results to $N$ subsystem, each of which has $m_i$ modes, and conclude that $\sum_{p'=1}^M\tilde\lambda_{pp'}(x) \mathcal B(x,p') = \sum_{i=1}^N\sum_{p'_i=1}^{m_i} \mu_i \tilde\lambda_{p_ip'_i}(x_i)\mathcal B_{i{p'_i}}(x_i)$ with $M = \Pi_{i=1}^Nm_i$. By applying the following inequality
	\begin{align}\label{Eq:22}
		\rho_{\mathrm{int}_{i_{p_i}}}(s_1+\cdots+s_{N-1})\leq\sum_{i=1}^{N-1}\rho_{\mathrm{int}_{i_{p_i}}}((N-1)s_i),
	\end{align}
	which is valid for any $\rho_{\mathrm{int}_{i_{p_i}}}\in\mathcal{K}_\infty\cup \{0\}$, and
	any $s_i\in\R_{\ge0}$, $i\in\{1,\cdots,N\}$, employing condition~\eqref{eq:LB_c1} and Assumption~\ref{Asu:1}, one can obtain the chain of inequalities in \eqref{Eq:23}. By defining
	\begin{align}\notag
		\kappa (s)  &\Let \min\Big\{-\mu^T(-\Lambda+\Delta)\Gamma (\bar{\mathcal B}(x))\,\big|\, \mu^T \bar{\mathcal B}(x)=s\Big\}, \\\notag
		\psi&\Let\sum_{i=1}^N\mu_i\max_{p_i\in P_i}\{\psi_{ip_i}\},
	\end{align}
	where $\bar{\mathcal B}(x)=[\mathcal B_{1{p_1}}(x_1);\ldots;\mathcal B_{N{p_N}}(x_N)]$, condition \eqref{eq:B3} is also satisfied. Then $\mathcal B$ is a CBF for $\Sigma$, which completes the proof.
\end{proof}

\begin{figure*}[h]
	\rule{\textwidth}{0.1pt}
	\begin{align}\notag
		\mathcal{L} &\mathcal B(x,p)+ \sum_{p'=1}^M\tilde\lambda_{pp'}(x) \mathcal B(x,p')= \mathcal{L}\sum_{i=1}^N\mu_i\mathcal B_{i{p_i}}(x_i) + \sum_{i=1}^N\sum_{p'_i=1}^{m_i} \mu_i \tilde\lambda_{p_ip'_i}(x_i)\mathcal B_{i{p'_i}}(x_i)\\\notag
		&= \sum_{i=1}^N\mu_i\Big(\mathcal{L}\mathcal B_{i{p_i}}(x_i) + \sum_{p'_i=1}^{m_i} \tilde\lambda_{p_ip'_i}(x_i)\mathcal B_{i{p'_i}}(x_i)\Big)\leq \sum_{i=1}^N\mu_i\Big(- \kappa_{ip_i} (\mathcal B_{i{p_i}}(x_i))+\rho_{\mathrm{int}_{ip_i}}(\Vert w_i \Vert^2) + \psi_{ip_i}\Big) \\\notag
		&   \leq \sum_{i=1}^N\mu_i\Big(- \kappa_{ip_i} (\mathcal B_{i{p_i}}(x_i))+\rho_{\mathrm{int}_{ip_i}}(\sum_{j=1,i\neq{j}}^N\Vert w_{ij}\Vert^2) + \psi_{ip_i}\Big) \\\notag
		& = \sum_{i=1}^N\mu_i\Big(- \kappa_{ip_i} (\mathcal B_{i{p_i}}(x_i))+\rho_{\mathrm{int}_{ip_i}}(\sum_{j=1,i\neq{j}}^N\Vert y_{ji}\Vert^2) + \psi_{ip_i}\Big)  \\\notag
		&\leq \sum_{i=1}^N\mu_i\Big(- \kappa_{ip_i} (\mathcal B_{i{p_i}}(x_i))+\sum_{j=1,i\neq{j}}^N\rho_{\mathrm{int}_{ip_i}}((N-1)\Vert y_{ji}\Vert^2) + \psi_{ip_i}\Big) \\\notag
		& \leq \sum_{i=1}^N\mu_i\Big(- \kappa_{ip_i} (\mathcal B_{i{p_i}}(x_i))+\sum_{j=1,i\neq{j}}^N\rho_{\mathrm{int}_{ip_i}}((N-1)\Vert h_{j}(x_j)\Vert^2) + \psi_{ip_i}\Big)\\\notag
		& \leq \sum_{i=1}^N\mu_i\Big(- \kappa_{ip_i} (\mathcal B_{i{p_i}}(x_i))+\sum_{j=1,i\neq{j}}^N\rho_{\mathrm{int}_{ip_i}}((N-1)\alpha_{jp_j}^{-1}(\mathcal B_{j{p_j}}( x_j))) + \psi_{ip_i}\Big) \\\notag
		& \leq \sum_{i=1}^N\mu_i\Big(-\hat\lambda_{ip_i}\hat\gamma_{i}(\mathcal B_{i{p_i}}( x_i))+\sum_{j=1,i\neq{j}}^N\hat\delta_{{ij}p_j}\hat\gamma_{j}(\mathcal B_{j{p_j}}(x_j)) + \psi_{ip_i}\Big) \\\label{Eq:23}
		&=\mu^\top(-\Lambda+\Delta)\Gamma(\mathcal B_{1{p_1}}(x_1);\ldots;\mathcal B_{N{p_N}}( x_N))+\sum_{i=1}^N\mu_i\psi_{ip_i}\leq - \kappa(\mathcal B(x,p)) + \psi.
	\end{align}
	\rule{\textwidth}{0.1pt}
\end{figure*}

\begin{proof}\textbf{(Theorem~\ref{th:sumprod})}
	For $\bar p \in \mathcal{AP}$, consider an accepting state run $\mathcal{R}^{\bar p}$ and $\mathcal{P}^{\bar p}(\mathsf{q})$ as the set of state runs of length $3$. For $\aleph=(q,q',q'') \in \mathcal{P}^{\bar p}(\mathsf{q})$, one can verify from Lemma~\ref{Lemma} that the upper bound on the probability that the solution process of $\Sigma$ starts at $X_0=\mathsf L^{-1}(\bar\sigma(q,q'))$ and reaches $X_u=\mathsf L^{-1}(\bar\sigma(q',q''))$ within the time horizon $[0,\mathcal T] \subseteq \mathbb R_{\ge 0}$ under the control input $\nu_{\aleph}$ is given by $\delta_{\aleph}$. Now the upper bound on the probability that the trace of the solution process reaches the accepting state following the path corresponding to $\mathsf{q}$ is given by the product of the probability bounds corresponding to all elements $\aleph=(q,q',q'') \in \mathcal{P}^{\bar p}(\mathsf{q})$:
	\begin{equation*}
	\mathds{P}\{\bar\sigma(\mathsf{q})\models \mathcal{A}^c \} \leq\prod_{\aleph\in\mathcal{P}^{\bar p}(\mathsf{q})}\{\delta_{\aleph} \,\big|\, \aleph=(q,q',q'') \in \mathcal{P}^{\bar p}(\mathsf{q})\}. 
	\end{equation*}
	Now one can conclude that the final upper bound on the probability for the solution process of $\Sigma$ starting from any initial state $\xi(0)=x_0\in \mathsf L^{-1}(\bar p)$ to violate the required specification is essentially the summation of probabilities of all possible accepting state runs of $\mathcal{A}^c$, \emph{i.e.},
	\begin{equation*}
	\mathds{P}_\varrho^{x_0}\{\bar\sigma_\xi \models \mathcal{A}^c\} \leq\sum_{q \in \mathcal{R}^{\bar p}} \prod_{\aleph\in\mathcal{P}^{\bar p}(\mathsf{q})}\{\delta_{\aleph} \,\big|\, \aleph=(q,q',q'')\in \mathcal{P}^{\bar p}(\mathsf{q})\}. 
	\end{equation*}	
\end{proof}

\begin{proof}\textbf{(Lemma~\ref{sos})}
	Since $\mathcal B_p(x)$ and $l_p(x)$ in~\eqref{eq:sos0} are sum-of-squares, we have $0\leq \mathcal B_{p}(x)- l_{p}^T(x) g(x)-\alpha_{p}(h(x)^Th(x))$. Since the term $l_{p}^T(x) g(x)$ is non-negative over $X$, the new condition~\eqref{eq:sos0} implies the condition~\eqref{eq:LB_c1} in Definition~\ref{eq:local barrier}. Similarly, we can show that \eqref{eq:sos1} and \eqref{eq:sos2} imply conditions \eqref{eq:LB_c2} and~\eqref{eq:LB_c3} in Definition~\ref{eq:local barrier}. Now we show that condition~\eqref{eq:sos3} implies~\eqref{eq:B_c4}, as well. By selecting external inputs $\nu_{j}=l_{\nu_{j_p}}(x)$ and since terms $\hat l_{p}^T(x,\nu,w) g(x), l_{\nu_{p}}^T(x,\nu,w) g_{\nu}(\nu), l_{w_{p}}^T(x,\nu,w) g_{w}(w)$ are non-negative over the set $X$, we have $\mathcal{L}\mathcal B_{p}(x) + \sum_{p'=1}^m\tilde\lambda_{pp'}(x)\mathcal B_{p'}(x)\leq - \kappa_{p} (\mathcal B_{p}(x)) +\rho_{\mathrm{int}{p}}(w^Tw)+\psi_{p}$ which implies that the function $\mathcal B_p(x)$ is a CPBF and completes the proof. 
\end{proof}

\end{document}